\PassOptionsToPackage{table,xcdraw}{xcolor}
\documentclass[acmsmall]{acmart}

\usepackage{hyperref}
\usepackage{url}
\usepackage{amsmath,amsfonts}
\usepackage{algorithmic}
\usepackage{graphicx}
\usepackage{textcomp}
\usepackage{xcolor}
\usepackage{booktabs}
\usepackage{todonotes}
\usepackage{enumitem}
\usepackage{multicol}
\usepackage{longtable}
\usepackage{multirow}
\usepackage{nicematrix}
\usepackage{pifont}
\newcommand{\cmark}{\ding{51}}%
\usepackage{subcaption}
\usepackage{adjustbox}
\usepackage{array}

\newcolumntype{R}[2]{%
    >{\adjustbox{angle=#1,lap=\width-(#2)}\bgroup}%
    l%
    <{\egroup}%
}

\def\rot{\rotatebox}

\usepackage{listings, xcolor}

\definecolor{verylightgray}{rgb}{.97,.97,.97}

\lstdefinelanguage{Solidity}{
	keywords=[1]{anonymous, assembly, assert, balance, break, call, callcode, case, catch, class, constant, continue, constructor, contract, debugger, default, delegatecall, delete, do, else, emit, event, experimental, export, external, false, finally, for, function, gas, if, implements, import, in, indexed, instanceof, interface, internal, is, length, library, log0, log1, log2, log3, log4, memory, modifier, new, payable, pragma, private, protected, public, pure, push, require, return, returns, revert, selfdestruct, send, solidity, storage, struct, suicide, super, switch, then, this, throw, transfer, true, try, typeof, unchecked, using, value, view, while, with, addmod, ecrecover, keccak256, mulmod, ripemd160, sha256, sha3}, 
	keywordstyle=[1]\color{blue}\bfseries,
	keywords=[2]{address, bool, byte, bytes, bytes1, bytes2, bytes3, bytes4, bytes5, bytes6, bytes7, bytes8, bytes9, bytes10, bytes11, bytes12, bytes13, bytes14, bytes15, bytes16, bytes17, bytes18, bytes19, bytes20, bytes21, bytes22, bytes23, bytes24, bytes25, bytes26, bytes27, bytes28, bytes29, bytes30, bytes31, bytes32, enum, int, int8, int16, int24, int32, int40, int48, int56, int64, int72, int80, int88, int96, int104, int112, int120, int128, int136, int144, int152, int160, int168, int176, int184, int192, int200, int208, int216, int224, int232, int240, int248, int256, mapping, string, uint, uint8, uint16, uint24, uint32, uint40, uint48, uint56, uint64, uint72, uint80, uint88, uint96, uint104, uint112, uint120, uint128, uint136, uint144, uint152, uint160, uint168, uint176, uint184, uint192, uint200, uint208, uint216, uint224, uint232, uint240, uint248, uint256, var, void, ether, finney, szabo, wei, days, hours, minutes, seconds, weeks, years},	
	keywordstyle=[2]\color{teal}\bfseries,
	keywords=[3]{block, blockhash, coinbase, difficulty, gaslimit, number, timestamp, msg, data, gas, sender, sig, value, now, tx, gasprice, origin},	
	keywordstyle=[3]\color{violet}\bfseries,
	identifierstyle=\color{black},
	sensitive=true,
	comment=[l]{//},
	morecomment=[s]{/*}{*/},
	commentstyle=\color{gray}\ttfamily,
	stringstyle=\color{red}\ttfamily,
	morestring=[b]',
	morestring=[b]"
}

\newcommand{\new}[1]{#1}

\begin{document}

\date{}


\title{Vulnerability Detection in Ethereum Smart Contracts via Machine Learning: A Qualitative Analysis}

\author{Dalila Ressi}
\email{dalila.ressi@unive.it}
\affiliation{%
  \institution{Ca' Foscari University}
  \city{Venice}
  \country{Italy}
}
\author{Alvise Spanò}
\email{alvise.spano@unive.it}
\affiliation{%
  \institution{Ca' Foscari University}
  \city{Venice}
  \country{Italy}
}

\author{Lorenzo Benetollo}
\email{lorenzo.benetollo@unive.it}
\affiliation{%
  \institution{University of Camerino}
\city{Camerino}
  \country{Italy}
}
\affiliation{%
  \institution{Ca' Foscari University}
\city{Venice}
  \country{Italy}
}

\author{Carla Piazza}
\affiliation{%
  \institution{University of Udine}
  \city{Udine}
  \country{Italy}}
\email{carla.piazza@uniud.it}

\author{Michele Bugliesi}
\email{bugliesi@unive.it}
\affiliation{%
  \institution{Ca' Foscari University}
  \city{Venice}
  \country{Italy}
}

\author{Sabina Rossi}
\email{sabina.rossi@unive.it}
\affiliation{%
  \institution{Ca' Foscari University}
  \city{Venice}
  \country{Italy}
}

\renewcommand{\shortauthors}{Ressi, et al.}

\ccsdesc[500]{General and reference~Surveys and overviews}
\ccsdesc[500]{Security and privacy~Vulnerability management}
\ccsdesc[500]{Computing methodologies~Machine learning}

\keywords{Vulnerability Detection, Machine Learning, Ethereum Smart Contracts}

\begin{abstract} 

Smart contracts are central to a myriad of critical blockchain applications, from financial transactions to supply chain management. However, their adoption is hindered by security vulnerabilities that can result in significant financial losses. 
Most vulnerability detection tools and methods available nowadays leverage either static analysis methods or machine learning. Unfortunately, as valuable as they are, both approaches suffer from limitations that make them only partially effective. In this survey, we analyze the state of the art in machine-learning vulnerability detection for Ethereum smart contracts, by categorizing existing tools and methodologies, evaluating them, and highlighting their limitations. Our critical assessment unveils issues such as restricted vulnerability coverage and dataset construction flaws, providing us with new metrics to overcome the difficulties that restrain a sound comparison of existing solutions. 
Driven by our findings, we discuss best practices to enhance the accuracy, scope, and efficiency of vulnerability detection in smart contracts. Our guidelines address the known flaws while at the same time opening new avenues for research and development.
By shedding light on current challenges and offering novel directions for improvement, we contribute to the advancement of secure smart contract development and blockchain technology as a whole.

\end{abstract}

\maketitle

\section{Introduction}

Blockchain technology has gained widespread adoption as an effective infrastructure for developing decentralized applications. Among the many platforms currently available, Ethereum has emerged as (one of) the most popular, the first to introduce smart contracts as executable programs that enforce the rules specified by their code. 


Like their companions from other platforms \cite{bartoletti2024smart}, Ethereum smart contracts are programmed to serve a variety of applications: insurance refunds, financial transactions, corporate operations, tracking of goods in supply chains, IP protection \cite{hu2021comprehensive,lin2022survey}. They are also used by Decentralized Autonomous Organizations (DAO) \cite{wang2019decentralized}, real estate transactions \cite{ullah2021conceptual, karamitsos2018design}, decentralized finance \cite{chen2020blockchain}, and  legal industry \cite{waltl2019blockchains}.
Specific to Ethereum is the underlying architecture and toolchain: Ethereum smart contracts are compiled into bytecode and  deployed onto the Ethereum blockchain, where they are executed by the Ethereum Virtual Machine (EVM) in a trustless, decentralized~environment.

As a relatively new and fast-growing technology,  smart contracts suffer from a wide range of security vulnerabilities that can be exploited by malicious actors \cite{atzei2017survey,vidal2024openscv}. 
Besides causing millions of dollars in damages, such attacks hinder the trustworthiness and reputation of not just Ethereum smart contracts but also blockchain technology in general. 
A relevant example is the "DAO Hack" Ethereum blockchain experienced in 2016. The Decentralized Autonomous Organization (DAO), a smart contract on the Ethereum blockchain designed as a venture capital fund, was exploited by a hacker who stole 3.6 million ETH (about \$50 million worth at the time). The contract presented a \textit{reentrancy} vulnerability, where the callee contract was called before the update of the balance, allowing the malicious contract to `re-enter' and perform withdraw operations in a loop (see the Appendix for an example of reentrancy in Solidity). 

This event divided the Ethereum community on how to respond. One group supported a hard fork to reverse the theft and return the stolen funds to investors, while another group opposed it, citing the principle of immutability in blockchain technology.
On July 20, 2016, Ethereum implemented a hard fork at block 1,920,000, resulting in two separate blockchains. Ethereum (ETH) adopted the hard fork, effectively undoing the DAO hack and continuing to develop with further updates, including the transition to Ethereum 2.0. It remains a leading blockchain platform, known for its smart contracts and decentralized applications.
Ethereum Classic (ETC), the original chain, rejected the hard fork and preserved the blockchain history, including the DAO hack. The ETC community emphasizes immutability, believing the blockchain should be an unalterable record of transactions. Despite having a smaller user base and developer community, Ethereum Classic continues to operate with its own ecosystem.
Even though this attack was made very famous due to its heavy consequences, many smart contracts unfortunately still exhibit the reentrancy vulnerability\footnote{hHystorical collection of reentrancy attacks available at \url{https://github.com/pcaversaccio/reentrancy-attacks}}, like in the attacks to  Rari Capital/Fei in 2022 \footnote{Explained at \url{https://www.linkedin.com/pulse/understanding-preventing-reentrancy-attacks-crypto-hashlock-aqifc}} or the latest CurveFinance in 2023 \footnote{Explained at \url{https://hacken.io/discover/curve-finance-liquidity-pools-hack-explained/}}.
Detecting and addressing these issues is critical to prevent attacks capable of inflicting substantial financial losses and undermining the credibility of blockchain technology.


Given the relevance of the problem, a plethora of solutions have been designed and proposed from both research institutions as well as corporations. 
Most of these tools and frameworks target Solidity~\footnote{Solidity website \url{https://soliditylang.org/}} source code
or else directly on the contracts' bytecode/opcode retrieved from the blockchain. 
Similar to what happens in other programming languages and platforms, static analysis and other formal methods are predominant in this field. However, machine learning (ML) based solutions are recently becoming more popular due to their speed and range of application~\cite{ressi2024ai}.
In the present work, we focus on ML-based techniques for vulnerability detection in Ethereum smart contracts written in Solidity. Developed in the wake of the quickly expanding popularity of Deep Learning techniques for Natural language Processing and, more recently, for generative language models, such frameworks are gaining widespread adoption and progressively emerging as superior to static analyzers when measuring their performance in terms of both precision and speed of the analysis.

\subsection{Contributions}

Several surveys on Ethereum smart contract vulnerability detection exist in the literature but mostly concentrate on static analyzers \cite{tolmach2021survey,ghaffarian2017software,chen2020survey}. Also, the few with a specific focus on ML frameworks appear to have a rather limited scope both in width (the number of frameworks analyzed) and in-depth (the metrics based on which the frameworks are measured comparatively).   
To our knowledge, only two surveys have been conducted on the topic of machine learning for vulnerability detection in smart contracts \cite{surucu2022survey,kiani2024ethereum}. The first survey covers research published up to 2021, while the second and most recent survey primarily offers a limited taxonomy of the machine learning methods employed, without delving into detailed insights about the efficacy of the models or the dataset construction method. Moreover, they do not provide a critical evaluation of the drawbacks and limitations of the examined works or discuss open problems.

By contrast, we identify 26 different detectors available and compare them according to multiple metrics. We also provide our insights on open problems and propose solutions and mitigation strategies that can guide toward many possible future works. Specifically, we provide: 
\begin{itemize}
    \item A comprehensive review of the state-of-the-art machine learning-based methods for language-based vulnerability detection in smart contracts;
    \item An extensive analysis of existing methods according to different aspects, such as the model, the dataset, and the vulnerabilities considered;
    \item A consistent mapping from all different definitions of the vulnerabilities across the revised papers to a common taxonomy;
    \item An in-depth analysis of drawbacks and gaps in examined methods;
    \item A discussion of possible solutions and mitigation strategies to contrast emerged problems.
\end{itemize}

\subsection{Structure}
\new{Our survey is structured as follows. In Section \ref{Sec:Methodology}, we outline the methodology used to select the papers and the research questions we aim to answer. Section~\ref{sec:Back} provides essential background information introducing key concepts such as the Ethereum blockchain, the smart contract vulnerabilities considered, principal machine learning and static analysis techniques, and the datasets used. In Section \ref{sec:ML}, we discuss various representations of smart contracts and present the machine learning-based detection methods we analyzed. Section~\ref{Sec:Findings} addresses the research questions, while Section \ref{sec:Problems} examines the problems and limitations of the collected methods. Finally, we present our conclusions in Section \ref{sec:Conclusion}.
}

\section{Methodology}\label{Sec:Methodology}

For our analysis, we considered articles published from 2018 up to now. We retrieved relevant articles from Scopus and Google Scholar resulting from the query "smart contracts" AND "vulnerability detection" AND ("machine learning" OR "AI" OR "deep learning"). We then excluded methods that did not imply AI techniques or did not target smart contracts written in Solidity. Given their high number of citations, we also included a few papers only available as Arxiv. Indeed, in the machine learning community, it is often the case that unpublished papers are recognized as valid by researchers,
even if not officially published. As a result, we collected 26 papers about AI or ML for vulnerability detection in Ethereum smart contracts written in Solidity.
\medskip

Other than simply analyzing the literature, we focus on the following research questions:
\begin{itemize}
    \item \textbf{RQ1}: what vulnerabilities are the most considered by ML-based detectors?
    \item \textbf{RQ2}: what datasets are adopted or designed for the training?
    \item \textbf{RQ3}: which ML models are used for the detection?
    \item \textbf{RQ4}: are the proposed techniques and models comparable in terms of accuracy?
\end{itemize}



\section{Background}
\label{sec:Back}

To understand our analysis of machine learning-based vulnerability detectors we provide a brief description of key concepts, including the Ethereum blockchain, the Ethereum Virtual Machine and the Solidity programming language.
In particular, we focus on the several vulnerabilities that may occur in contracts, providing some examples with brief code snippets in the Appendix.
Next, an introduction to the most significant machine learning models and techniques allows the reader to better understand how machine learning-based vulnerability detectors work.
Finally, we introduce vulnerability detectors based on formal methods, as they are either exploited by ML for the dataset construction or the ML model is compared against them.
Our goal is to provide the reader with all the knowledge needed to compare the methods we considered in this survey. Specifically, we present core concepts about the Ethereum blockchain, vulnerabilities, principal machine learning models, and static analysis methods and datasets exploited in the papers we selected.

\subsection{The Ethereum Blockchain}

Among the many blockchain technologies available today, Ethereum, a pioneering platform introduced in 2015, stands out for its programmable nature and widespread adoption.
The Turing completeness provided by the platform allows developers to express complex logic, but this flexibility comes with challenges in ensuring secure coding practices.
The evolution of smart contracts has seen notable accidents where vulnerabilities were exploited, leading to financial losses and security concerns. 
The infamous "DAO Hack" in 2016 exploited a flaw in the smart contract code, resulting in the theft of a significant amount of Ether. 
Another attack, the "Parity Wallet Bug" in 2017 \cite{destefanis2018smart}, demonstrated the impact of vulnerabilities in smart contract wallets.
Such episodes underscore the importance of thorough security audits and the need for robust coding practices in smart contract development.
As the blockchain ecosystem continues to grow, understanding and addressing smart contract vulnerabilities is crucial for ensuring the reliability and security of decentralized applications.

Smart contracts on the Ethereum blockchain \cite{wood2014ethereum} are programs running on the platform that perform currency transfers, payments, and transactions in general.
Written in high-level languages like Solidity, these contracts are executed on the Ethereum Virtual Machine (EVM), providing a trustless and transparent way to interact with decentralized applications (DApps) and blockchain-based systems.
The source code is compiled into Ethereum-specific intermediate code, represented by opcodes that the EVM understands.
The compiled code, known as bytecode, is a series of bytes that the EVM can execute. 
It represents the machine-readable version of the source code.
The EVM is the runtime environment on the Ethereum network responsible for executing smart contracts.
It interprets and executes the bytecode, ensuring the deterministic execution of contracts across the decentralized network.

In the Ethereum ecosystem, an \emph{account} is identified by an Ethereum address and holds Ether (ETH) balances.
On the other hand, a \emph{contract address} is specifically associated with a smart contract and uniquely identifies it.
Transactions sent to a contract address trigger the execution of the smart contract's code, altering its state or producing output.
Platforms like Etherscan\footnote{\url{https://etherscan.io}.} facilitate the exploration of the Ethereum blockchain. 
Users can input a contract address to access details such as the contract's source code, bytecode, transaction history, and associated information. 
Etherscan enhances transparency, allowing users to verify and understand the behavior of smart contracts deployed on the Ethereum blockchain.

\subsection{Smart Contracts Vulnerabilities}
\label{sec:vulnerabilities}

A smart contract is a program written in a language specifically tailored to the target blockchain that typically runs on the platform's virtual machine.
The programming language mostly adopted to implement smart contracts for Ethereum is Solidity.
Since smart contracts are stored in the blockchain, immutability ensures that once a smart contract has been deployed it cannot be modified.
While this mechanism provides a certain level of protection against malicious entities and guarantees the fairness of the contract \cite{malakhov21}, it also makes it impossible to fix any bug that might be discovered in the contract's code.
Consequently, it is essential to provide developers with an array of tools to create secure smart contracts and rigorously test their code for potential security flaws \textit{before} their deployment.

In order to detect and address such vulnerabilities, various studies propose a way to categorize and define them. 
The first proposal belongs to Atzei et al.~\cite{atzei2017survey}, where the authors identify 12 vulnerabilities, and group them according to whether they can be related to the Solidity language, the EVM, or the underlying blockchain platform.
From this point, many other authors proposed a taxonomy for vulnerabilities in Ethereum smart contracts, for example by grouping them according to whether they can be exploited by attacks (security) or if they go against the intended behavior of the program (functionality) \cite{arganaraz2020detection}. 
Other works prioritize the level of threat the vulnerabilities pose and group them accordingly \cite{tsankov2018securify}. 
Over time, the number of detected problems steadily increased, up to 54 \cite{rameder2022review} and 94 in one of the most recent works \cite{vidal2024openscv}.
Specifically, Vidal et al. \cite{vidal2024openscv} recently proposed a new taxonomy, \new{namely OpenSCV}, where they identify a very large number of smart contracts vulnerabilities, often referred to with different names or abbreviations, and categorize them into 7 different categories. 
\new{The Decentralized Application Security Project top 10 (DASP TOP 10)\footnote{https://dasp.co} is a taxonomy often adopted by vulnerability detectors but, as the name suggests, it only reports ten of the most common vulnerabilities. Other commonly adopted classifications include the Common Weakness Enumeration (CWE)\footnote{\url{https://cwe.mitre.org/about/index.html}} and the Smart Contract Weakness Classification (SWC)\footnote{\url{http://swcregistry.io/}}.
Out of all the taxonomies and categorizations in the literature, OpenSCV}
can be considered the most rigorous to date, as it collects the largest amount of known vulnerabilities while indexing all the different names that have been used in literature for the same type of vulnerability.

\definecolor{high}{rgb}{0.99, 0.8, 0.8}
\newcommand{\High}{\cellcolor{high} High}
\definecolor{avg}{rgb}{0.99, 0.99, 0.8}
\newcommand{\Avg}{\cellcolor{avg} Avg}
\definecolor{low}{rgb}{0.8, 0.99, 0.8}
\newcommand{\Low}{\cellcolor{low} Low}

\begin{table}[th!]
\centering
\fontsize{6}{6}\selectfont
\setlength{\tabcolsep}{2pt} 
\renewcommand{\arraystretch}{1.6} 
\resizebox{0.95\textwidth}{!}{

\begin{NiceTabular}{cllccccl}[hvlines]
\textbf{Category} & \textbf{Vulnerability} & \textbf{OpenSCV} & \textbf{CWE} & \textbf{SWC} & \textbf{Acronym} & \textbf{Severity} & \textbf{Fixed}
\\
\Block{6-1}{Unsafe External Calls} 
    & Reentrancy                    & 1.1       & CWE-841 & SWC-107     & RE & \High & \\
    & Malicious Fallback Function   & 1.2       & CWE-685 & -           & MFF & \High &  \\
    & Unchecked External Call       & 1.3.1     & CWE-252 & SWC-104     & UEC & \Low    & EIP-150 2016 \\ 
    & Unchecked Low-level Call      & 1.3.3     & CWE-372 & -           & ULC & \Low    & v0.8.20 \\  
    & Version Compatibility Issues  & 1.5       & CWE-843 & -           & VCI & \Low &  \\  
    & Dangerous Delegate Call       & 1.6       & CWE-829 & SWC-112     & DDC & \High &  \\

\Block{3-1}{Mishandled Events} 
    & Mishandled Exception          & 2.1.1     & CWE-248 & -           & ME & \Avg &  \\ 
    & DoS by Gas Limit              & 2.1.2     & CWE-400 & SWC-128     & GL & \Avg &  \\ 
    & Missing Thrown Exception      & 2.2.1     & CWE-474 & -           & MTE & \Low &  \\ 

\Block{1-1}{Gas Depletion} 
    & Gas Consumption               & 3.1       & CWE-691 & SWC-126     & GC & \Avg & \\

\Block{2-1}{Erroneous Credit Transfer} 
    & Multiple Sends                & 4.2       & CWE-400 & SWC-105     & MS & \High & \\
    & Use of \texttt{send} 
      instead of \texttt{transfer}  & 4.3       & CWE-840 & -           & ST & \Avg & \\

\Block{11-1}{Bad Programming Practices \\ \& Language Weaknesses} 
    & Bad Randomness                        & 5.1       & CWE-330 & SWC-120     & BR & \High &  \\  
    & Assert Violation                      & 5.4.2     & CWE-670 & SWC-110     & AV & \Avg &  \\ 
    & Incorrect ERC20 Interface Override    & 5.6.1     & CWE-694 & -           & IIO & \Avg & v0.8.20 \\ 
    & Parameter Type Mismatch               & 5.6.3     & CWE-704 & -           & PTM & \Low & v0.8.20 \\  
    & Locked Ether                          & 5.7.1     & CWE-561 & -           & LE & \Avg & v0.8.20 \\    
    & Floating Pragmas                      & 5.8.1     & CWE-664 & SWC-103     & PRA & \Low  &   \\ 
    & Missing Constant Modifier             & 5.10.2    & CWE-701 & -           & MCM & \Low &  \\
    & Implicit Visibility                   & 5.10.3    & CWE-710 & SWC-108     & IV & \Low &  \\  
    & Variable Shadowing                    & 5.12.1    & CWE-1109 & SWC-119    & VS & \Avg & v0.6.0 \\
    & Unsafe Recast                         & -         & - & -                 & UR & \Avg &   \\    
    & Infinite Loop                         & -         & - & -                 & INL & \Low &   \\ 

\Block{4-1}{Incorrect Control Flow} 
    & Timestamp Dependency                  & 6.1.1     & CWE-829 & SWC-116     & TD & \Avg & \\ 
    & Transaction Order Dependency          & 6.1.2     & CWE-362 & SWC-114     & TOD & \High & \\ 
    & Improper Locking                      & 6.1.3     & CWE-667 & SWC-132     & IL & \Avg & \\ 
    & Short Address Attack                  & 6.2.1     & CWE-20  & -           & SAA & \Avg  &  \\  
 
\Block{2-1}{Arithmetic Issues} 
    & Integer Underflow                     & 7.1.1     & CWE-191 & SWC-101     & UF & \Avg &  v0.8.0 \\
    & Integer Overflow                      & 7.1.2     & CWE-190 & SWC-101     & OF &  \Avg & v0.8.0 \\

\Block{3-1}{Improper Access Control} 
    & Use of \texttt{tx.origin}             & 8.1.1     & CWE-1126 & SWC-115    & TXO & \High &   \\
    & Unprotected Ownership                 & 8.1.2     & CWE-732 &             & UO & \High & \\
    & Self-Destruct                         & 8.1.3     & CWE-1032 & SWC-106    & SD & \High & \\

\end{NiceTabular}}
\vspace{1em}
\caption{The table shows vulnerabilities in Solidity grouped into 8 categories \new{similar} to the OpenSCV classification \cite{vidal2024openscv}. We enlisted the vulnerabilities detected by multiple tools and fit them into those categories. The \emph{Severity} column shows the impact of the vulnerability on real-world scenarios. The Fixed column reports whether a specific vulnerability has been solved through a language or compiler fix.}

\label{tab:risks}
\end{table}

\medskip
To introduce the vulnerabilities mentioned in the ML detectors we examined, we mostly adopt Vidal's classification, which defines the 7 macro-groups as shown in Table \ref{tab:risks}, and we explain each of the sub-categories mentioned in the ML papers. This table is one of the main contributions of our work, as it is then exploited to re-map all differently named vulnerabilities across the papers examined to a common taxonomy. 
We use the OpenSCV nomenclature, or the name most commonly reported in the literature, or a new name that we believe represents the vulnerability well for each vulnerability. 
We provide the section number corresponding to the OpenSCV paper chapter discussing the vulnerability, as well as the code identifier for CWE and SWC.
We also assign an acronym to facilitate referencing throughout this paper. Finally, we report the severity related to each vulnerability and if it has been proposed a fix so far.

One of the flaws in \new{most categorizations} is the lack of a risk assessment associated with these vulnerabilities. Since it is vital for a vulnerability detector to address the most critical \new{issues}, \new{we} associate them with three different \new{severity levels: high, medium, and low. These levels are partially inherited from \cite{tsankov2018securify,arganaraz2020detection} and partially reconstructed according to the authors' experience.} 
It is interesting to point out how systematically categorizing these vulnerabilities can be influenced by the experience of different developers. The task is further complicated by some vulnerabilities fitting into multiple macro-categories. For instance, Unchecked External Call (UEC) and Unchecked Low-Level Call (ULC) are similar, differing only in whether the function called is an external contract or a local assembly routine.
Moreover, ULC and Dangerous Delegate Calls (DDC) as defined in the papers are sometimes used for the same vulnerability. 
In other cases, some vulnerabilities like the call-stack depth attack \cite{luu2016making} can fall under different categories such as ULC and Mishandled Exception (ME).
Although our taxonomy is not exhaustive with respect to all existing vulnerabilities and covers only a subset of the vulnerabilities described by Vidal et al., it remains critically important. By mapping different nomenclatures of vulnerabilities found in the machine learning papers we analyzed to the same system, we make it easier for researchers working on new models to reference related work.


\newcommand{\catstyle}[1]{\texttt{\small{\textbf{#1}}}}
\newcommand{\catother}[1]{\emph{#1}}
\newcommand{\cat}[1]{\hfill \\}

\begin{description}[leftmargin=1.5em]
  
\item[Unsafe External Calls] 
    \cat{Reentrancy, Malicious Fallback Function, Unchecked External Call, Unchecked Low-level Call, Version Compatibility Issues, Dangerous Delegate Call}
    These vulnerabilities arise when external calls are not properly managed, potentially leading to unintended behaviors and security exploits in the smart contract's execution.
    \catstyle{Reentrancy} occurs when a contract makes an external call to another contract before completing its state changes.
    An attacker can exploit this by repeatedly entering the target contract, potentially leading to unintended behavior or unauthorized access to the contract's state.
    \catstyle{Malicious Fallback Function} is a generalization of the mechanism by which the reentrancy attack is achieved.
    An attacker can deploy a contract with a fallback function that contains malicious logic. 
    This can include actions such as reverting the transaction or consuming excessive gas.
    \catstyle{Unchecked External Call} is sometimes referred to as \catother{Unchecked Return Value} \cite{fu2019critical} or \catother{No Check After Contract Invocation} \cite{chen2020soda} and consists of a call to an external contract that has failed silently and whose return value indicating success or failure is not tested by the caller code.
    This leads to programs assuming an operation succeeded when it failed, eventually yielding to inconsistent states.
    This is also found under the name \catother{Call Stack Depth} in the literature, with \citet{luu2016making} classifying it as a \catother{Mishandled Exception}, whereas \citet{vidal2024openscv} calls it \catother{Unchecked Return Value}.
    We prefer the latter classification as this vulnerability happens when the call stack is over due to too many external contract invocations, though the EVM does not raise an exception but rather returns false from the call that exceeds the stack limit.
    This vulnerability was fixed in 2016 with the Ethereum Improvement Proposal 150 \footnote{An Ethereum Improvement Proposal, or EIP for short, is an official ticketing system for proposing changes to the Solidity language or the Ethereum platform in general.}.
    \catstyle{Unchecked Low-level Calls} involves using low-level call functions without proper checks and validation.
    It can result in unintended consequences, such as sending Ether to an unintended address or executing arbitrary code.
    When mentioned as \catother{Inline Assembly} or \catother{Check Effects} \cite{zhang2022smart}, it refers to the risks associated with using inline assembly code within smart contracts. 
    Solidity allows developers to include assembly code directly within their contracts using the \lstinline{assembly} keyword.
    While this practice provides flexibility and allows for low-level interactions with the EVM, it also introduces potential security vulnerabilities if not used carefully.
    \catstyle{Version Compatibility Issues} relates to interoperability issues between contracts built in different language versions.
    Newer contracts may execute or inherit discontinued functionality present in older contracts.
    In the literature, it may appear under the name \catother{Assembly Usage} \cite{tsankov2018securify}.
    \catstyle{Dangerous Delegate Calls} occur when a contract uses the \lstinline{delegatecall} primitive to invoke code from another contract.
    If not handled carefully, delegate calls can lead to unexpected behavior, including the unintended modification of the calling contract state.
    \\

\item[Mishandled Events] 
    \cat{Mishandled Exception, DoS by Gas Limit, Missing Thrown Exception}
    This category encompasses issues related to events that are not properly handled by the code, such as exception handling and other control-flow-breaking mechanisms.
    This can lead to unintended consequences, potential exploitation, and disruptions in the normal operation of the smart contract, making it vulnerable to DoS attacks.
    \catstyle{Mishandled Exception} refers to uncaught exceptions or other cases of wrong exception handling, which can lead to unwanted behaviors or unexpected errors.
    This is often mentioned as \catother{Unhandled Exception} \cite{tsankov2018securify}. 
    The \catstyle{DoS by Gas Limit} vulnerability is a denial-of-service that exploits the gas limit mechanism in the Ethereum network to disrupt or hinder the execution of a smart contract.
    This vulnerability typically affects functions within a smart contract that iterate over a dynamic array or perform loops that may consume an unpredictable amount of gas.
    If the gas required to complete these operations exceeds the gas limit, the transaction will fail.
    This vulnerability appears in the literature under a multitude of different names that could be misleading at times, such as \catother{Call in Loop} \cite{tsankov2018securify}, \catother{DoS with Unbounded Operation} \cite{kaleem2020vyper,nassirzadeh2022gas} or \catother{Unbounded Mass Operation} \cite{grech2020madmax}.
    The \catstyle{Missing Thrown Exception} vulnerability occurs when a smart contract fails to include appropriate error-handling mechanisms, such as throwing exceptions or reverting transactions when an error condition is met.
    This is also known under the name \catother{ERC20 Transfer} \cite{ashizawa2021eth2vec} or \catother{Missing the Transfer Event} \cite{chen2020soda}.
    \\

\item[Gas Depletion] 
    \cat{Gas Consumption}
    This category includes vulnerabilities related to gas exploitation and limits. 
    Gas is the unit of computation on the Ethereum network and every transaction has a computational limit.
    Attacks involve intentionally crafting a smart contract or transaction that consumes an excessive amount of gas.
    The \catstyle{Gas Consumption} vulnerability leads to transactions failing due to reaching the gas limit, making the contract unusable or susceptible to DoS attacks.
    Exceeding the gas limit can cause the entire transaction to fail and the loss of the associated gas.
    Managing gas consumption effectively is crucial for maintaining the functionality and security of Ethereum smart contracts.
    This vulnerability is also known by the name \catother{}{Opaque Predicate} \cite{chen2020gaschecker}.
    \\

\item[Erroneous Credit Transfer] 
    \cat{Multiple Sends, Use of send instead of transfer}
    This category encompasses severe vulnerabilities that involve the wrong management of transactions.
    \catstyle{Multiple sends} occur when a smart contract initiates multiple external transactions or sends within a single function call.
    Each external transaction consumes gas, which might result in exceeding the gas limit. In this case, all the sends will fail, and if that function is the only way to move ether out of the contract, the smart contract will stay in a stuck state.
    In the same context, a denial-of-service can happen even without the presence of multiple sends, for example, if an expensive operation is called or if a function is applied to an arbitrarily large collection.
    This vulnerability is also known as \catother{Ether Leak} \cite{choi2021smartian} and \catother{Unchecked Transfer Value}~\cite{zhuang2020smart}.    
    The \catstyle{Use of send instead of transfer} issue happens when these functions are used for transferring Ether. 
    The \lstinline{send} and \lstinline{transfer} primitives both allow the execution of a credit transfer.
    However, in the case of a problem, \lstinline{transfer} will abort the process with an exception, whereas \lstinline{send} function will return false, and transaction execution is continued. 
    An attacker may manipulate the \lstinline{send} function and be able to continue executing a credit transfer operation without proper authorization.
    This vulnerability may be considered a special case of the \catother{Unchecked External Call} category.
    \\

\item[Bad Programming Practices \& Language Weaknesses] 
    \cat{Bad Randomness, Assert Violation, Incorrect ERC20 Interface Override, Parameter Type Mismatch, Locked Ether, Floating Pragmas, Missing Constant Modifier, Implicit Visibility, Variable Shadowing, Unsafe Recast, Infinite Loop}
    This is a broad category including several bad coding practices as well as language weaknesses that lead to unpredictable behavior.
    \catstyle{Bad Randomness} - sometimes called \catother{Predictable Variable} \cite{zhou2019devign,zhang2022multi} - refers to the use of insecure methods for generating random numbers within smart contracts. 
    Since Ethereum smart contracts run in a deterministic environment, creating truly random numbers is challenging. 
    Inadequate randomness can be exploited by malicious actors to predict outcomes and manipulate the results of functions that depend on randomness, such as lotteries, games, or any contract requiring randomness for fairness.
    Common bad practices are using \lstinline{block.timestamp} as a source of randomness, or hashing predictable data through \lstinline{keccak256}.
    \catstyle{Assert Violation} may occur when a malicious caller deliberately invokes a function passing arguments that would not succeed the pre-conditions checks.
    These might be exploited by attackers to manipulate the contract's state or trigger unintended consequences.
    Proper handling and validation of conditions are crucial to prevent assert violations and ensure the security and correctness of the smart contract.
    OpenSCV calls this vulnerability \catother{Wrong Selection of Guard Function} \cite{vidal2024openscv}.    
    The \catstyle{Incorrect ERC-20 Interface Override} arises when a smart contract incorrectly implements or overrides the standard ERC20 interface functions. 
    Common mistakes such as returning the incorrect values or writing an incorrect function signature, can lead to unexpected behavior, security vulnerabilities, and interoperability issues with other contracts and services that rely on the ERC20 standard.
    This is also known under the short name \catother{ERC20 Interface} \cite{tsankov2018securify} or \catother{Missing Return Statement} \cite{hu2023detect}.
    The \catstyle{Parameter Type Mismatch} vulnerability occurs when a smart contract function is called with parameters that do not match the expected types or order of types defined in the function signature. 
    If there is a mismatch, the function may not execute as intended, and in some cases, this can be exploited.
    This is also known in the literature under the name \catother{ERC20 Indexed} \cite{tsankov2018securify}.
    \catstyle{Locked Ether}, or \catother{Locked money} \cite{feist2019slither}, refers to a situation where Ether, the native cryptocurrency of the Ethereum blockchain, becomes permanently stuck or inaccessible within a smart contract. This can occur due to design flaws in the smart contract's code.
    \catstyle{Floating Pragmas} refers to the risks associated with using complex or conditional pragmas in the code.
    Pragmas are special compiler directives that provide instructions or settings for the Solidity compiler.
    While pragmas can be useful for setting compiler options or enabling/disabling certain features, using complex or conditional pragmas can introduce potential security vulnerabilities.
    This is also known as \catother{Solc Version} \cite{tsankov2018securify}.
    The \catstyle{Missing Constant Modifier} vulnerability (a.k.a. \catother{Constable States} \cite{tsankov2018securify}) refers to situations where a function that does not modify the state is not marked with the appropriate visibility and non-mutating keywords (\lstinline{view} or \lstinline{pure}).
    This can lead to inefficiencies and increased gas costs, as the EVM might treat the function as stateful, which incurs higher costs.
    \catstyle{Implicit Visibility} refers to the default visibility qualifier for state variables being \lstinline{internal} in Solidity.
    Forgetting to specify a visibility qualifier, such as \lstinline{private}, may leave the door open to unwanted calls by external agents.
    \catstyle{Variable Shadowing} consists of the definition of a local variable having the same name as a state variable, factually shadowing it.
    The last two vulnerabilities are not enumerated within the OpenSCV classification.
    \catstyle{Unsafe Recast} refers to converting data types into one another without proper validation or type checking, for example converting a big number into a shorter integer word. 
    This is a common source of bugs and vulnerabilities in contracts and can result in security issues such as integer overflow/underflow, memory corruption, or incorrect data manipulation.
    \catstyle{Infinite Loop} in smart contracts can lead to the contract getting stuck in a code loop without completing its functions.
    This can be exploited by attackers to consume excessive gas, leading to a DoS situation.
    In OpenSCV this is wrongly associated with the \catother{Extraneous Exception Handling} vulnerability \cite{vidal2024openscv}, though \citet{liu2021smart} considers it a different kind of issue.
    \\

\item[Incorrect Control Flow]
    \cat{Timestamp Dependency, Transaction Order Dependency, Improper Locking, Short Address}
    The category includes various risks related to the flow of control within the contract's execution.
    \catstyle{Timestamp Dependency} Contracts that rely on the timestamp of a block for certain decisions are vulnerable to timestamp dependency.
    Attackers can manipulate the timestamp to their advantage, potentially affecting the outcome of the contract's logic.
    Mitigating these vulnerabilities involves implementing robust control flow mechanisms, avoiding dependencies on specific transaction orders or timestamps, and ensuring proper input validation to prevent short-address attacks.
    Another name for this vulnerability is \catother{Block Information} \cite{li2022eosioanalyzer} and \catother{Block Info Dependency} \cite{chen2022wasai}, which happens when a smart contract uses block information (like \lstinline{block.number}) for logic and decision-making.
    However, miners can manipulate some of these details to a small extent, potentially leading to vulnerabilities if the contract overly relies on them for critical logic.
    \catstyle{Transaction Order Dependency} arises when a smart contract relies on the order in which transactions are mined within a block.
    If the contract logic depends on specific transaction ordering, it can be exploited by miners to impact the outcome of the program by manipulating the order of transactions.
    The \catstyle{Improper Locking} vulnerability involves issues with the use of mutexes (mutual exclusions) to prevent reentrancy attacks or ensure that only one instance of a function is executing at any given time.
    Improper implementation of locking mechanisms can lead to vulnerabilities, including race conditions and reentrancy attacks.
    This is also called \catother{Strict Check for Balance} \cite{chen2020soda} and \catother{Arbitrary Sending of Ether} \cite{feist2019slither}.
    A \catstyle{Short Address Attacks} occurs when a contract does not properly check the length of input data during an external call.
    An attacker can exploit this by sending data with insufficient length, causing the contract to misinterpret the input and potentially leading to unexpected behavior.
    This vulnerability is also known as \catother{Invalid Input Data} \cite{chen2020soda} and \catother{Unchecked Input Arguments} \cite{li2022vulnerability}.
    \\

\item[Arithmetic Issues] 
    \cat{Integer Underflow, Integer Overflow}
    This category involves issues related to wrong manipulation of arithmetic data types in computations.
    \catstyle{Integer Underflow} occurs when the result of an arithmetic operation is less than the minimum representable value, which could lead to balances becoming much larger than expected.
    Attackers may exploit underflow vulnerabilities to manipulate balances or quantities in a way that benefits them.
    Mitigating these vulnerabilities involves using safe arithmetic libraries, data types, or checks to prevent overflow and underflow.  
    \catstyle{Integer Overflow} occurs when the result of an arithmetic operation exceeds the maximum representable value for the data type used.
    In smart contracts, this can lead to unexpected behavior, as the overflowed value might wrap around to a smaller value than expected.
    Exploiting integer overflow vulnerabilities can allow attackers to manipulate balances, quantities, or other numeric values within a contract.
    Both integer overflows and underflows have been addressed in v0.8.0 of Solidity.
    \\

\item[Improper Access Control]
    \cat{Tx.origin, Unprotected Ownership, Self-destruct}
    This category groups vulnerabilities that arise when the permissions and access rights within a contract are not correctly managed. 
    The \catstyle{Tx.origin} vulnerability refers to the potential security risk that arises when a smart contract uses the \lstinline{tx.origin} global variable for authentication. 
    This can lead to scenarios where an attacker tricks the contract into executing actions that it shouldn't, by exploiting the difference between \lstinline{tx.origin} and \lstinline{msg.sender}.
    It is also known in the literature as \catother{Incorrect Check for Authorization} \cite{chen2020soda} or \catother{Missing Authorization Verification} \cite{chen2022wasai}.
    The \catstyle{Unprotected Ownership} vulnerability in Solidity refers to the lack of proper access control mechanisms to restrict the execution of certain critical functions to only the contract owner or authorized entities. 
    This vulnerability can lead to unauthorized users gaining control over the contract and its assets, resulting in significant security risks.
    Many different names have been given to this, among which \catother{Tainted Owner Variable} \cite{brent2020ethainter}, \catother{Freeze Account} \cite{ma2023hgat} and \catother{Vulnerable Access Control} \cite{zhuang2020smart}.
    In our classification, the name \catstyle{Self-Destruct} underpins a number of vulnerable scenarios that go under different names in the literature.
    In Ethereum, a programmer can plan the self-destruction of a smart contract invoking the \lstinline{selfdestruct} primitive.
    This destroys the contract and deletes its code from the blockchain, sending its remaining Ether to a specified address.
    However, if not properly managed, this feature can be exploited.
    Sometimes referred to as \catother{Guard Suicide} \cite{chang2019scompile}, \catother{Unprotected Suicide} \cite{hu2023detect,mavridou2019verisolid} or simply \catother{Suicide} \cite{fu2019critical}, this happens when an attacker deliberately destroys a smart contract, eventually performing an Ether transfer to a specific smart contract that was not supposed to receive it.
    Notably, \catother{Accessible self-destruct} and \catother{Tainted self-destruct} \cite{brent2020ethainter} are special cases, respectively where the self-destruct mechanism is too easily accessible (likely due to poor access controls or design flaws) or where the attacker can set to which address the remaining balance of the smart contract is sent.

\end{description}

\subsection{Machine Learning Techniques}
Machine Learning (ML) has transformed computational data processing by utilizing data for autonomous predictions and pattern discovery. Within this domain, Supervised Learning (SL) and Unsupervised Learning (UL) are foundational, guiding models through labeled and unlabeled datasets. SL focuses on classification tasks, instructing models to label new data based on known examples, while UL aims to extract cohesive data groups without pre-labeled information.

Classic algorithms such as K-Nearest Neighbors (kNN), Decision Trees (DT), Support Vector Machine (SVM), and Logistic Regression (LR) have offered initial solutions to diverse problems. Ensemble methods like Random Forest (RF) combine multiple models for improved performance, mitigating overfitting and enhancing generalization.
Despite some blurring of distinctions, many refer to these techniques simply as ML, while Artificial Intelligence (AI) often involves neural networks.
Neural Networks (NN) have led to Deep Learning (DL), where networks automatically extract intricate data representations \cite{dargan2020survey}. DL's depth varies based on the complexity of the task, typically determined by the number of hidden layers in the architecture. Various neural network architectures exist, tailored to different data types and problem domains. Convolutional Neural Networks (CNN) specialize in spatial data, while Recurrent Neural Networks (RNN), including variants like Long Short-Term Memory (LSTM) and Gated Recurrent Unit (GRU), manage temporal dependencies, crucial for sequence tasks \cite{alzubaidi2021review}. 
In particular, two mechanisms recently showed to improve the performance of these architectures, namely the attention mechanism and the design of bidirectional modules.
The attention mechanism enhances context-aware processing, particularly important for sequence-related tasks. 
It is also particularly meaningful in the context of explainability, as it explains what features the model learns as relevant to the task.

The second technique consists of connecting two hidden layers of opposite directions to the same output. In this way, it is possible to link past and future inputs providing the network with more context.
Sequential data are also the core of Natural Language Processing (NLP) \cite{khurana2023natural}, which broadens ML's reach, enabling computers to understand and generate human language. 
NLP has a wide range of applications, like language recognition, text analysis, and translation. A particular type of NPL is Large Language Models (LLM), large networks trained with vast amounts of text that can learn statistical distributions and can be used for many tasks. One of the first examples is BERT (Bidirectional Encoder Representations from Transformers), which has transformed language understanding by capturing word context bidirectionally \cite{devlin2018bert}. LLMs are also closely related to generative models, like Generative Adversarial Networks (GANs) and autoencoders, which create realistic data samples \cite{makhzani2015adversarial}. 
The most notable case is OpenAI ChatGPT, which can generate not only text but also audio and images in real-time.

Finally, another type of architecture is gaining popularity. It is the case of Graph Neural Networks (GNNs), which excel in tasks involving graph-structured data, making them valuable for social network analysis and recommendation systems \cite{wu2022graph}.
In the context of vulnerability detection in Ethereum smart contracts, many of these models have been utilized. Detection tasks often involve supervised classification, where models are trained on labeled contracts for binary or multi-class prediction.

\subsection{Static Analysis Techniques}\label{sec:formal}


Static analysis has long been recognized as an effective method for security verification, and recent years have seen the development of various frameworks dedicated to assessing the security of Ethereum smart contracts \cite{di2019survey,ghaleb2020effective,kushwaha2022ethereum}.
Among the methodologies employed by these frameworks are abstract interpretation, symbolic execution, and taint analysis. These methodologies detect potential vulnerabilities by examining the code without executing it, allowing for the identification of potential security flaws before deployment.
These static analysis techniques are implemented in numerous tools that have been prominently featured in the literature and are heavily exploited in the creation and development of datasets intended for training Machine Learning algorithms to detect vulnerabilities in Ethereum smart contracts. This significance prompts their inclusion in our discussion.

Typically, static analysis tools work by examining the Solidity source code or a disassembled version of the compiled contract, whether in bytecode or opcode form. These tools identify potential errors or vulnerabilities, providing an essential layer of inspection. 
Below, we provide a brief overview of the most commonly used static analysis techniques for vulnerability detection in Ethereum contracts, along with a compilation of tools that implement these techniques and are relevant for the application of machine learning algorithms.
One widely used method is \textit{control flow analysis} \cite{denning77}, which inspects the sequence of execution paths within the contract to uncover irregularities such as unreachable code or infinite loops. Another essential technique is \textit{data flow analysis} \cite{fosdick1976}, which tracks the flow of data through the contract to identify vulnerabilities like uninitialized storage variables or improper handling of user inputs. \textit{Symbolic execution} \cite{symbolic76} is also popular; it systematically explores possible execution paths by treating inputs as symbolic values, thereby detecting issues like integer overflows, reentrancy attacks, and other logic errors. Additionally, \textit{taint analysis} \cite{taint2018} helps in identifying potentially malicious data flows by marking inputs and observing how they propagate through the contract.

There are many other formal techniques that are currently used for this task and hundreds of different tools and frameworks are available in the literature. Here we mention the tools used by the machine learning detectors we take into consideration. 

Introduced in 2018, \textbf{Slither} \cite{feist2019slither} employs data flow analysis on SlithIR, an intermediate representation of Solidity code. This versatile tool excels at pinpointing a variety of bugs and vulnerabilities, including shadowing, uninitialized variables, reentrancy issues, locked ether, and arbitrary ether transfers. Beyond security analysis, Slither also aids in code comprehension, and automated optimization detection, and streamlines the code review process.
Dating back to 2017, \textbf{Mythril} \cite{mythril} is a robust analysis tool that harnesses symbolic execution, SMT solving, and taint analysis to scrutinize Ethereum smart contracts. Beyond Ethereum, it extends its capabilities to other blockchain platforms as well. Mythril is proficient at uncovering vulnerabilities and weaknesses in contract logic, providing valuable insights into contract security.
\textbf{Oyente} \cite{luu2016making} leverages symbolic execution to analyze a spectrum of critical properties, including transaction ordering, timestamp dependency, code reentrancy, and exception handling. Its capabilities extend to identifying intricate vulnerabilities related to these properties, making it a valuable addition to the Ethereum smart contract security toolkit. This tool is no longer maintained (as of 2024). 
\textbf{Honeybadger} \cite{torres2019art} is an analysis tool to detect honeypots based on Oyente.
Released in 2018, \textbf{Securify} \cite{tsankov2018securify} also employs symbolic execution but takes as input the contract bytecode and a predefined set of security patterns. This approach allows Securify to detect vulnerabilities by systematically exploring the contract's behavior, offering comprehensive coverage and accuracy in vulnerability detection.

     \textbf{Vandal} \cite{brent2018vandal} is a platform released in 2018 designed to detect potential security vulnerabilities in compiled contract bytecode. It allows for quick development and prototyping of new vulnerability specifications written in Datalog. With Vandal, one can create any program analysis over the intermediate representation of a contract. It comes with a static program analysis library that includes many useful Datalog relations. This tool is no longer maintained (as of 2024).
     \textbf{Osiris} \cite{torres2018osiris} is an analysis tool to detect integer bugs in Ethereum smart contracts released in 2018. It employs a combination of two approaches, which are symbolic execution and taint analysis. Osiris is based on Oyente. This tool is no longer maintained (as of 2024).
     \textbf{SmartCheck} \cite{tikhomirov2018smartcheck} is an extensible static analysis tool for discovering vulnerabilities and other code issues in Ethereum smart contracts written in Solidity. It translates Solidity source code into an XML-based intermediate representation and checks it against XPath patterns. It detects approximately 20 types of vulnerabilities like implicit visibility level, compiler version not fixed, arithmetic division, style guide violation, etc. This tool is no longer maintained (as of 2024).
     \textbf{Dedaub} \cite{smaragdakis2021symbolic} is a Security Suite that provides a set of tools for the analysis of Smart Contracts, among which is one of static analysis. They refer to the static analysis technique under the name of symbolic analysis, which stands for symbolic+value-flow static analysis.


     \textbf{Manticore} \cite{mossberg2019manticore} is a symbolic execution tool released in 2017 for the analysis of smart contracts and binaries. It can analyze Ethereum smart contracts (EVM bytecode), Linux ELF binaries, and WASM Modules. It can find reentrancy and underflow/overflow vulnerabilities. This tool is now maintained by the community (as of 2024).
     \textbf{Solhint}\footnote{The official Solhint website is at \url{https://protofire.io/projects/solhint}.} is an open-source project for linting Solidity code released in 2017. This project provides both Security and Style Guide validations. It can find different vulnerabilities like reentrancy and perform other useful checks.
     Although \textbf{RemixIDE}\footnote{The official RemixIDE website is at \url{https://remix-project.org/}.} is primarily an online integrated development environment (IDE), it also supports the static analysis of Smart Contracts. This capability allows users to leverage plug-ins to initiate analyses using Remix's analyzer, or through external tools such as Solhint or Slither.

While most existing frameworks predominantly rely on static analysis, a subset incorporates dynamic analysis or even combines both approaches to enhance precision. A notable exemplar in this category is \textbf{SmartScan\cite{samreen2021smartscan}}, a cutting-edge tool developed in 2021. SmartScan offers a fusion of static and dynamic analysis techniques tailored explicitly for identifying Denial-of-Service (DoS) attacks in Ethereum smart contracts. This amalgamation of methods equips SmartScan to deliver a high level of accuracy in detecting this particular class of vulnerabilities, showcasing the evolving landscape of Ethereum smart contract security tools.



\subsection{Datasets Available}

Given the public nature of the Ethereum blockchain, and also its popularity, it is not difficult to recover millions of smart contracts that have already been deployed. On the other side, the labeling process usually requires large amounts of time, especially if it is done manually. Having a large labeled dataset is useful not only for the training of ML models but also for assessing the performance of other detectors, such as those using formal methods. Machine learning-based methods for vulnerability detection either create their new dataset or exploit one of the following ones available.

\begin{figure}
    \centering
    \includegraphics[width=0.6\textwidth]{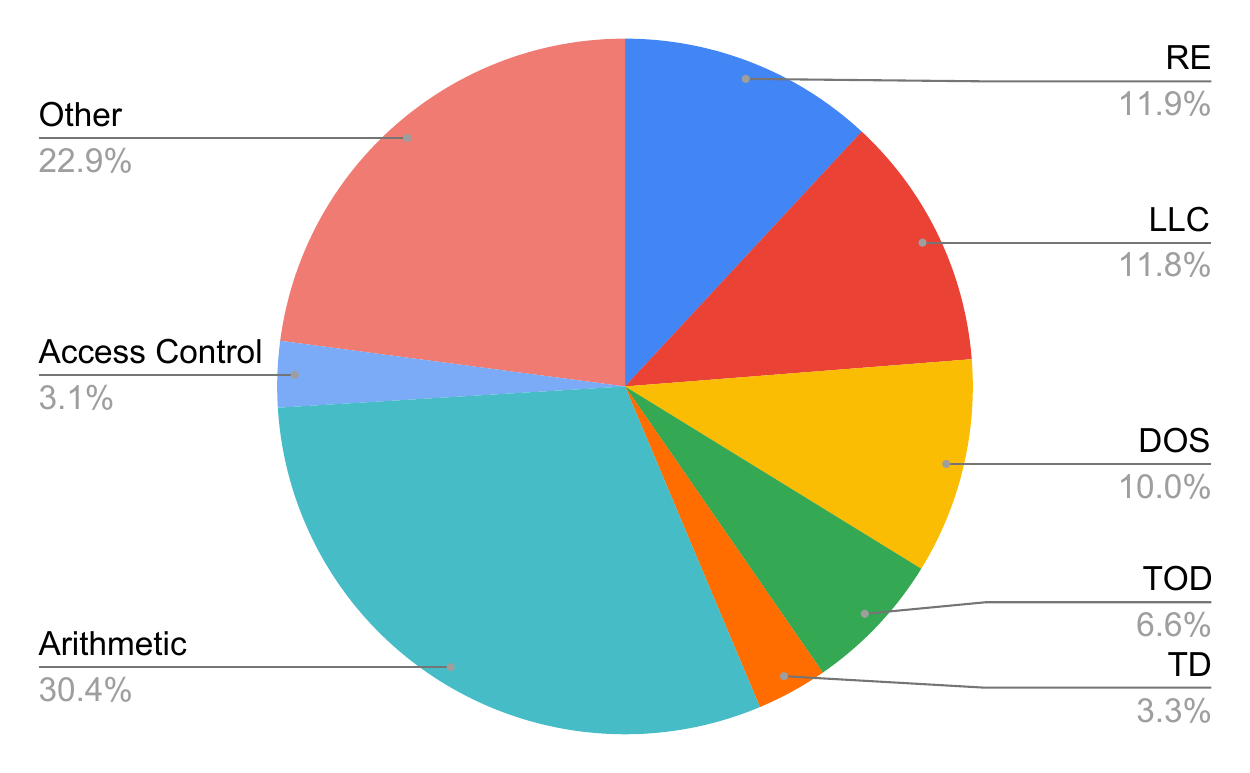}
    \caption{Frequency and distribution of vulnerabilities in Smartbugs Wild, detected by 9 different static analyzers [Source: Smartbugs Wild at https://github.com/smartbugs/smartbugs-results].}
    \label{fig:hist_smart}
\end{figure}

\textbf{CodeSmell} \cite{chen2020defining} is a dataset of manually labeled smart contracts, where the authors exploited not only real-world smart contracts but also smart-contract-related posts from Ethereum StackExchange. It consists of 587 real-world smart contracts presenting 20 different contract vulnerabilities. 
\textbf{SolidiFI}~\cite{ghaleb2020effective} exploits bug injection to introduce vulnerabilities in smart contracts in a controlled way. It works by adding pre-defined bug patterns at all possible locations in the AST (Abstract Syntax Tree) of smart contracts. It consists of 50 contracts injected by 9369 distinct bugs (Re-entrancy, Timestamp dep, Unchecked send, Unhandled exp, TOD, Integer flow, tx.origin, Miscellaneous). In the paper, the authors use it to evaluate six static analysis tools, Oyente, Securify, Mythril, SmartCheck, Manticore and Slither.
\textbf{Smartbugs} \cite{ferreira2020smartbugs,diAngeloEtAl2023ASE} is a scalable framework supporting different static analysis tools for vulnerability detection in Solidity contracts. It comes with two datasets, a small one made of 143 annotated vulnerable contracts with 208 tagged vulnerabilities (\textbf{Smartbugs Curated}), and a large one containing 47518 unique contracts collected through Etherscan (\textbf{Smartbugs Wild}). The authors used the dataset to test up to 19 different static analysis tools, which they used to label the larger dataset. The distribution of the vulnerabilities in SmartBugs Wild according to the authors can be seen in Figure \ref{fig:hist_smart}.
\textbf{HuangGai} \cite{huanggui} is an Ethereum smart contract bug injection framework that can inject 20 types of bugs into Solidity smart contracts. 
The authors release three datasets of 964, 4744, and 66205 contracts. The first two are injected with bugs that have been manually inspected, while the last one is made of real contracts.
\textbf{ESC} is a dataset of 40932 smart contracts presenting reentrancy, timestamp dependence, and infinite loop vulnerabilities. The labeling method has not been disclosed, and it is available thanks to \cite{yu2021deescvhunter}.

\section{Smart Contracts Vulnerability Detection with Machine Learning}\label{sec:ML}

Machine learning has already been proven to work very well for vulnerability detection not only in smart contracts but also in many other programming languages, such as Java, SQL, PHP, C/C++, and so on \cite{lin2020software,zeng2020software,wu2017vulnerability}. 
In the realm of Ethereum smart contract vulnerability detection, machine learning emerges as a potent solution, offering real-time detection across a broad spectrum of vulnerabilities. 
Effective data representation ensures that the relevant features of smart contracts are accurately captured and utilized, which directly influences the accuracy and efficiency of the detection process. Poor representation can lead to models being trained on incomplete or irrelevant information, reducing their ability to generalize across different vulnerabilities and contract variations.
Choosing suitable model architectures is also crucial for maintaining high accuracy and generalization.

\subsection{\new{Smart Contract Representations}}

As smart contract source code can be considered high-level and human-understandable, vulnerabilities can be detected by processing the input with techniques similar to NLP. On the other side, this is not the case if we are examining only the opcode or bytecode of such compiled contracts. As smart contracts cannot be fed as they are to the ML techniques or networks, the information that is extracted from the data is extremely important for the detection process. Data processing techniques come from various fields, not only NLP but also Information Retrieval (IR) \cite{zhang2022survey}. Most of them are generally used for text classification. 
There are many different types of data pre-processing, depending on how we want to represent the features critical for the detection phase \cite{galke2021bag,li2021review}. 
Here we present the most common ones, used not only for smart contracts but also for representing the source code of other programming languages with the purpose of vulnerability detection.

A common approach consists of deploying techniques to transform the code into a \textit{graph-based representation}.
The most used techniques include Abstract Syntax Tree (AST), Control Flow Graph (CFG), Data-Flow Graph (DFG), Program Dependence Graph (PDG), or a combination of them \cite{sabbaghi2020systematic}. These methods preserve the semantics of the contract and are often used in literature not only for vulnerability detection but also for other tasks concerning the security of the system. This graph representation is then either flattened to be fed to a neural network, or used as it is in combination with a GNN model.

Other techniques can be associated with the classic \textit{Bag-Of-Words (BOW)} model, where we lose the order of the words, or in this case the order of the code, to rather inspect the distribution of the contract \cite{hacohen2020influence}. The idea behind this concept is that vulnerable contracts should have similar distributions and thus we can detect new vulnerable contracts using how they are collocated within this learned distribution. We consider in this group pre-processing techniques such as term frequency-inverse document frequency (TF-IDF), word embeddings and one-hot vectors.

\new{Finally, some networks can take the code or its compiled version as it is. We can consider as belonging to this}
\textit{sequence-representation} order-preserving techniques such as n-grams, word2vec and even transforming the contract into an image \cite{chakraborty2021deep}.
Notice that these pre-processing techniques are often combined, and there is no hard boundary between them. The same concept applies to the machine learning models used for the detection. While we can group the methods according to the type of architecture, some papers deploy more than one architecture or combine two different ones.

\subsection{\new{Machine Learning Models}}

The most wide-spread approach consists in transforming the contracts in a Control Flow Graph (CFG). CFG is a graphic representation of control flow or of all paths that might be traversed through the execution of programs and it is often used in static analysis (e.g. Vandal). Specifically, \citet{mi2023automated} represent the opcodes of smart contracts in a CFG, and then they transform them into vectors with a depth-first search algorithm. From these vectors, they construct the feature matrix by using n-gram and term frequency-inverse document frequency (TF-IDF), where smart contracts are represented by vectors of numerical values. Finally, a Deep Neural Network (DNN) with custom loss is used as a binary classifier to discriminate vulnerable contracts. The authors test their method on two other different datasets, but they do not specify which vulnerability they can detect, as their method is a simple binary classifier. A similar technique is proposed in \new{HGAT} \cite{ma2023hgat}. First, using Abstract Syntax Tree (AST) and Control Flow Graph, the functions in the smart contract are abstracted into code graphs (CFG). Then the authors abstract each node in the code subgraph, extract the node features and by using the graph attention mechanism GAT, splice the obtained vectors to form the features of each line of statements. They use a particular type of graph neural network called Hierarchical Graph Attention Network (HGAT) to detect 4 types of different vulnerabilities. However, the authors do not explain what the vulnerabilities consist of, as they simply exploit the labels provided by Smartbugs Wild, and the dataset they consider seems to include only vulnerable contracts.

One of the most popular ML models is Graph Neural Networks. Since GNNs are designed to work on data represented by graphs, they are particularly useful when we want to preserve the structure and the interactions taking place within a contract, as well as its semantics. Indeed, GNNs have been proven to be particularly effective in the literature for vulnerability detection in general \cite{zhou2019devign}.
In particular, \citet{cai2023combine} claims that using NLP performs worse than GNN when a smart contract has more complex logic and structure information. They propose a bidirectional gated graph neural network with a hybrid attention pooling layer to learn the code features, efficiently capturing vulnerability-related features from the graph for vulnerability detection. The model they propose is a binary classifier trained on 5 different datasets and presenting 9 different types of vulnerabilities. 

A similar graph representation data pre-processing technique is Data Flow Graph (DFG), widely used for program analysis. In particular, \citet{wu2021peculiar} claim that, unlike AST, DFG is the same under different abstract grammars for the same source code and that it provides crucial code semantic information for code understanding. They propose a detector called Peculiar to find reentrancy vulnerabilities on manually labeled smart contracts. Peculiar features a subgraph representation of DFG where they extract critical information, combined with GraphCodeBERT \cite{guo2020graphcodebert} a GNN pre-trained on a large dataset of six programming languages \cite{husain2019codesearchnet}.
Other detectors also deploy a GNN \cite{zhuang2020smart,liu2021combining}. 
In particular, Zhuang et al.~\cite{zhuang2020smart} \new{claim to} construct a contract graph to represent both syntactic and semantic structures of a smart contract function. They propose a degree-free graph convolutional neural network (DR-GCN) and a novel temporal message propagation network (TMP) to detect 3 types of vulnerabilities (reentrancy, timestamp dependence, and infinite loop). 
\new{The same authors} \cite{zhuang2020smart} later provided an improved version of their detector they called \new{CGE} \cite{liu2021combining}. The proposed method consists of extracting a graph representation of the contract where the nodes are labeled as core nodes, normal nodes, and fallback nodes, while the edges represent temporal dependencies. At the same time, they use a Feedforward Neural Network (FNN) to extract security patterns. Both of these representations are then fed to a GNN for vulnerability detection.

Using a graph representation of smart contracts is a popular solution because it allows to capture the semantics, and not only the syntax. In particular, the approach proposed in \cite{cai2024fine} combines both by exploiting AST and graph-structured semantic features, and they use a child-sum tree-LSTM cell to learn these heterogeneous features. Their approach is one of the few performing a statement-level detection, where they can identify which line of code presents the vulnerability. The authors provide code examples for each vulnerability they consider, while also defining \textit{security best practices} to avoid such issues. The construction of the dataset also contains the same contract where the authors replace safe code with vulnerable one.

Eth2Vec \cite{ashizawa2021eth2vec} also exploits different pre-processing techniques to preserve information, using AST as well as opcodes and bytecode representations. Then, an EVM Extractor analyzes EVM bytecodes syntactically and creates JSON files for instruction level, block level, function level, and contract level. The authors use an unsupervised approach using a model that takes JSON files generated from bytecodes as input and then computes the code similarity for each contract.


Another approach combines unsupervised learning and graph neural networks \cite{huang2021hunting}. The authors propose an unsupervised graph embedding algorithm to vectorize the code graphs of normalized slices and measure the vector similarities for bytecode matching and vulnerability detection. The normalization allows the comparison of bytecode resulting from different versions of the compiler. The potentially vulnerable smart contracts can be identified by measuring the similarities between their vectors and known vulnerable ones.
Another tool for detecting unknown vulnerabilities is proposed in \cite{zhang2022smart}. The authors extract custom features, called Crucial Operation Sequence (COS), which are then given to a pre-trained BERT model, followed by a multi-objective-based NN for detection (MODNN). The model achieves superior performance with respect to other methods \cite{ashizawa2021eth2vec, liao2019soliaudit, zhuang2020smart} while being more robust against new types of vulnerabilities. 
The authors refrain from describing the vulnerabilities, opting instead to build the dataset using contracts from Smartbugs Wild. These contracts are then labeled according to the outputs from Smartcheck, Osiris, and Mythril.
A less recent work \cite{liao2019soliaudit} also focused on finding unknown vulnerabilities. The authors evaluated multiple ML methods on an impressive span of 13 different vulnerabilities, where using n-gram with TF-IDF and logistic regression achieved about 90\% F1-score. By using the dynamic contract fuzzing technique, they claim to be able to find also unknown vulnerabilities by measuring the Minkowski distance on the smart contracts transactions. This method tests the accuracy of many different machine learning techniques, finding that the best combination consists of using n-grams and logistic regression. 

Other authors focused on detecting unknown vulnerabilities by training a classifier to understand if a contract exhibits correct behavior or not
\cite{gu2022detecting,li2022detecting}.  This approach allows a higher degree of scalability than the classical detectors, as it allows one to detect vulnerabilities that have not been seen so far, but it has a lower accuracy with respect to the others. The difference between the two proposals lies in the structure of the model. In particular, \citet{li2022detecting} first deploys a binary classifier that divides normal and abnormal opcodes, followed by a second classifier that distinguishes between known and unknown vulnerabilities. Differently, \new{CNN-BiLSTM} \cite{gu2022detecting} cannot separate known and unknown vulnerabilities in the opcodes, as it is based on the concept that vulnerable opcodes have similarities to the opcode sequences of transactions containing known vulnerabilities.

Other methods try a wide range of ML techniques \cite{zhang2022cbgru,wang2020contractward}. \citet{zhang2022cbgru} combine two different embedding techniques (word2vec and FastText) with three different models (CNN, GRU and LSTM), where for the last two they also test the bidirectional version. As a result, they propose a hybrid model (CBGRU) that achieves better results with respect to the single models previously proposed. The authors use Smartbugs Wild for the experiments but report an unrelated paper for the labeling method they used.
Similarly, \citet{jain2023integrated} use Word2vec pre-processing together with embedding with transformer and bi-GRU (close to the model proposed in \cite{cai2023combine}), while for the classification they exploit a simple CNN. They achieve a remarkable 96.5\% F1-score on average and compare with many other ML models (SVM, LSTM, GNN) and detectors (see Table \ref{tab:methods}). The authors perform program slicing, which removes lines of code that are not relevant for the detection, while instead preserving the code containing sensitive operations (\lstinline{delegatecall} or \lstinline{call}) and sensitive data (variables that can be controlled by miners, such as \lstinline{block.number} or \lstinline{block.timestamp}). The work includes an ablation study to show that combining AST, CFG, PDG, and using slicing and bidirectional graphs achieve superior performance. Even if the authors also exploit techniques to overcome their very unbalanced dataset, such as SMOTE and Tomek, they do not explicitly provide the number of contracts per class they consider, which for a couple of classes seems to not be greater than 10. 

ContractWard \cite{wang2020contractward} also exploits the SMOTETomek API for balancing the training set. The method consists of extracting bigram features from simplified opcodes and testing the accuracy of multiple machine learning-based approaches using XGBoost for training the models. The authors classified functionally similar opcodes into one category to reduce the dimensionality of the features. For example, e.g., PUSH1, PUSH32 and so on, are classified into a PUSH opcode. They showed their method can predict six different types of vulnerabilities with an F1-score over 96\%. A very similar approach was already proposed by the \citet{song2019efficient}, where they adopted a One vs. Rest (OvR) algorithm for multi-label classification of six vulnerabilities on 50K verified smart contracts from Ethereum's official website. Their method exploits bigram for feature extraction and simplifies opcodes by dislodging the operands and classifying functionally similar opcodes into one category. The authors also deployed SMOTE for balancing underrepresented classes. Using only simple ML methods such as RF, SVM and KNN, the authors report outstanding results, with F1-scores higher than 90\% in all classes, with RF achieving the highest accuracy.

Other than \cite{jain2023integrated}, another work adopts a transformer architecture for the detection \cite{jeon2021smartcondetect}. By using BERT architecture the authors return a vulnerability report as a JSON file containing filename, vulnerability rules, and corresponding functions, inspired by the definition provided in \cite{tikhomirov2018smartcheck}.

Another commonly used architecture for smart contracts vulnerability detection is LSTM, such as in \cite{qian2020towards, yu2021deescvhunter}. \citet{qian2020towards} transform source code into code snippets, containing only lines of code that are relevant for the detection. After manual labeling, the lines are further transformed first into tokens and then into vectors. Such representation is then fed to a bidirectional LSTM with an attention mechanism \new{(BLSTM-ATT)}.
\new{Combining LSTM with the attention mechanism seems to be one of the best solutions, as} reported also by \cite{yu2021deescvhunter}. The authors propose \new{DeeSCVHunter}, a custom pre-processing method using a vulnerability candidate slice (VCS) which contains rich syntax characteristics and semantic information. They try different architectures (RNN, GRU LSTM, CNN-LSTM, TextCNN) and report that a Bi-LSTM using VCS outperforms other methods in reentrancy detection (86.87\% F1-score) while the combination of LSTM with attention and VCS works the best for timestamp dependency (79.93\% F1-score).

The attention mechanism is exploited in many of the proposed methods \cite{ma2023hgat,cai2023combine,zhang2022spcbig,sun2021attention}.
\citet{zhang2022spcbig} (who are the same group from \cite{zhang2022cbgru}) propose \new{SPCBIG-EC}, an ensemble of models including GRU and CNN, which automatically pick the best detector according to the examined vulnerability thanks to the attention mechanism. \new{Similarly to their other work, they study a hybrid method, and they also design a serial-parallel convolutional layer (SPC) for feature extraction.} They achieve high F1-scores on a balanced dataset, consisting of more than 1300 per each of the six vulnerabilities considered, at the expense of a high inference time (about 10 seconds) per contract, due to the time needed to run all the models. The method proposed in \cite{sun2021attention}, instead, adopts a simpler approach, consisting of pre-processing the opcode into slices normalizing them and then applying Word2vec. The authors exploit a CNN with self-attention \new{(ABCNN)} for the detection phase.




\new{A very different technique} consists of transforming the source code of smart contracts (or related bytecodes) into 2D images, which are then fed to a classical Convolutional Neural Network (CNN). 
Both \cite{huang2018hunting} and \cite{hwang2022codenet} convert contracts’ bytecode to RGB colors and then use the CNN model to detect vulnerabilities on a manually labeled dataset. Codenet \cite{hwang2022codenet} preserves spatial information by using 1-D convolutions without stride.

Finally, machine learning can be combined with other verification techniques, such as fuzzing~\cite{grieco2020echidna}. In particular, xFuzz \cite{xue2022xfuzz} exploits ML to significantly reduce the search space for exploitable paths. The authors focus only on reentrancy vulnerability, but they are able to detect cross-contract attacks when the number of contracts is greater than two.

\begin{table*}[!ht]
    \centering
    \fontsize{9}{9}\selectfont
    \setlength{\tabcolsep}{2pt} 
    \renewcommand{\arraystretch}{1.2} 
    \resizebox{\textwidth}{!}{
    \begin{tabular}{|ll|p{4.8cm}|rllllr|lllll|llllllllll|}
    \cline{4-24}
    \multicolumn{2}{c}{}& \multicolumn{1}{c}{ }& \multicolumn{6}{|c|}{Dataset statistics} & \multicolumn{5}{c|}{Dataset used}& \multicolumn{10}{c|}{Labeling}\\
    \hline
        Ref & Name & Vulnerabilities & \rot{90}{\# Vulnerabilities} & \rot{90}{Are Vulnerability Explained? } & \rot{90}{Has \# Contracts per Class?} & \rot{90}{Is Dataset Balanced?} & \rot{90}{Uses Smote/Tomek?} & \rot{90}{\# Contracts}  & \rot{90}{SmartBugs Wild} & \rot{90}{SolidiFi} & \rot{90}{ESC} & \rot{90}{Other} & \rot{90}{Custom} & \rot{90}{SecuriFI} & \rot{90}{Smartcheck} & \rot{90}{Vandal} & \rot{90}{Dedaub} & \rot{90}{Osiris} & \rot{90}{Solhint} & \rot{90}{Remix} & \rot{90}{Manual} & \rot{90}{Other} & \rot{90}{Not specified} \\ \hline
        
        \cite{cai2024fine} & - & OF ULC SD RE TOD GL UR & 7 & \cmark &  & ~ & ~ & 6000 & ~ & ~ & ~ & ~ &\cmark& ~ & ~ & ~ & ~ & ~ & ~ & ~ & ~ & \cmark & ~ \\ \hline
        \cite{ma2023hgat} & HGAT & RE TD UF OF & 4 & & \cmark &\cmark& ~ & 7018 &\cmark& ~ & ~ & ~ & ~ & ~ & ~ & ~ & ~ & ~ & ~ & ~ & ~ & ~ & ~ \\ \hline
        \cite{jain2023integrated} & - & RE TXO UF OF UEC ME TD TOD ULC ULC BR ULC SD & 13 & &  &\cmark&\cmark& n.a. &\cmark& ~ & ~ &\cmark& ~ & ~ & ~ & ~ & ~ & ~ & ~ & ~ & ~ & ~ & ~ \\ \hline
        \cite{mi2023automated} & - & - & - & &  & ~ & ~ & 160012+587+9369 & ~ &\cmark& ~ &\cmark&\cmark& ~ & ~ & \cmark & ~ & ~ & ~ & ~ & ~ & ~ & ~ \\ \hline
        \cite{cai2023combine} & Bi-GGNN & TD DDC OF RE SD SAA TD TOD UEC & - & \cmark &  & ~ & ~ & 9369+964+143+7000 &\cmark&\cmark& ~ &\cmark& ~ & ~ & ~ & ~ & ~ & ~ & ~ & ~ & ~ & ~ & ~ \\ \hline
        \cite{zhang2022smart} & MODNN & UF OF UEC TOD TD RE AV TXO ULC ULC TD ULC & 12 & & \cmark & ~ & ~ & 18796 &\cmark&\cmark& ~ &\cmark&\cmark& ~ & \cmark & ~ & ~ & \cmark & ~ & ~ & ~ & ~ & ~ \\ \hline
        \cite{zhang2022cbgru} & CBGRU & INL RE OF UEC TD UF & 6 & & \cmark &\cmark& ~ & >10000 &\cmark& ~ & ~ & ~ & ~ & ~ & ~ & ~ & ~ & ~ & ~ & ~ & ~ & ~ & ~ \\ \hline
        \cite{zhang2022spcbig} & SPCBIG-EC & RE TD INL OF UF UEC & 6 & & \cmark &\cmark&\cmark& >10000 & ~ & ~ & ~ & ~ &\cmark& \cmark & ~ & ~ & ~ & ~ & ~ & ~ & ~ & ~ & ~ \\ \hline
        \cite{li2022detecting} & - & RE MFF UEC MTE IL TD TXO & - & & \cmark & ~ & ~ & 3328 & ~ & ~ & ~ & ~ &\cmark& ~ & ~ & ~ & ~ & ~ & ~ & ~ & ~ & ~ & \cmark \\ \hline
        \cite{xue2022xfuzz} & xFuzz & RE TXO DDC & 3 & \cmark & \cmark & ~ & ~ & 7391  & ~ & \cmark & ~ & ~ &\cmark& \cmark & ~ & ~ & ~ & ~ & \cmark & ~ & ~ & ~ & ~ \\ \hline
        \cite{hwang2022codenet} & CodeNet & RE ULC TD TXO & 4 & & \cmark & \cmark & ~ & 13443 & \cmark & \cmark & ~ &\cmark& ~ & \cmark & ~ & ~ & ~ & ~ & ~ & ~ & ~ & ~ & ~ \\ \hline
        \cite{gu2022detecting} & CNN-BiLSTM & TXO UEC MTE IL TD & 5 & & \cmark & ~ &  & 3262 & ~ & ~ & ~ & ~ &\cmark& ~ & ~ & ~ & ~ & ~ & ~ & ~ & ~ & ~ & \cmark \\ \hline
        \cite{ashizawa2021eth2vec} & Eth2Vec & RE TD MTE GC IV OF UF & 7 & &  & ~ & ~ & 95152 & ~ & ~ & ~ & ~ &\cmark& ~ & \cmark & ~ & ~ & ~ & ~ & ~ & ~ & ~ & ~ \\ \hline
        \cite{jeon2021smartcondetect} & SmartConDetect &  - & 23 & &  & ~ & ~ & 10000 & ~ & ~ & ~ & ~ &\cmark& ~ & \cmark & ~ & ~ & ~ & ~ & ~ & ~ & ~ & ~ \\ \hline
        \cite{sun2021attention} & ABCNN & RE OF UF TD & 4 & &  & ~ & ~ & 8632 & ~ & ~ & ~ &\cmark& ~ & ~ & ~ & ~ & ~ & ~ & ~ & ~ & ~ & ~ & ~ \\ \hline
        \cite{yu2021deescvhunter} & DeeSCVHunter & RE TD & 2 & &  & ~ & ~ & 40932 & ~ & ~ &\cmark& ~ & ~ & ~ & ~ & ~ & ~ & ~ & ~ & ~ & ~ & ~ & \cmark \\ \hline
        \cite{lutz2021escort} & ESCORT & UEC RE MS SD GL TOD AV & 7 & & \cmark & \cmark & ~ & 93497 & ~ & ~ & ~ & ~ &\cmark& ~ & ~ & ~ & \cmark & ~ & ~ & ~ & ~ & ~ & ~ \\ \hline
        \cite{wu2021peculiar} & Peculiar & RE & 1 & \cmark & \cmark & ~ & ~ & 1648 &\cmark& ~ & ~ & ~ &\cmark& ~ & ~ & ~ & ~ & ~ & ~ & ~ & \cmark & ~ & ~ \\ \hline
        \cite{huang2021hunting} & - & OF RE BR UO ME & 5 & \cmark & \cmark & ~ & ~ & 2297058+32499+24 & ~ & ~ & ~ & ~ &\cmark& ~ & ~ & ~ & ~ & ~ & ~ & ~ & ~ & \cmark & ~ \\ \hline
        \cite{liu2021combining} & CGE & RE TD INL & 3 &\cmark &  & ~ & ~ & 40932+4170 & ~ & ~ &\cmark&\cmark& ~ & ~ & ~ & ~ & ~ & ~ & ~ & ~ & ~ & ~ & \cmark \\ \hline
        \cite{zhuang2020smart} & TMP, DR-GCN & RE TD INL & 3 & \cmark &  & ~ & ~ & 40932+4170 & ~ & ~ &\cmark&\cmark& ~ & ~ & ~ & ~ & ~ & ~ & ~ & ~ & ~ & ~ & \cmark \\ \hline
        \cite{qian2020towards} & BLSTM-ATT & RE & 1 & \cmark & \cmark &\cmark& ~ & 4000 & ~ & ~ & ~ & ~ &\cmark& ~ & ~ & ~ & ~ & ~ & ~ & ~ & ~ & \cmark & ~ \\ \hline
        \cite{wang2020contractward} & ContractWard & OF UF TOD UEC TD RE & 6 & \cmark & \cmark &\cmark& ~ & 49502 & ~ & ~ & ~ & ~ &\cmark& ~ & ~ & ~ & ~ & ~ & ~ & ~ & ~ & ~ & ~ \\ \hline
        \cite{liao2019soliaudit} & Soliaudit & UF OF UEC TOD TD RE AV TXO ULC ULC TD ST SD & 13 & \cmark & \cmark &\cmark& ~ & 17979 & ~ & ~ & ~ & ~ &\cmark& ~ & ~ & ~ & ~ & ~ & ~ & \cmark & ~ & ~ & ~ \\ \hline
        \cite{momeni2019machine} & - & UF RE SD LE IIO MS MFF TXO UEC ULC ME VS VCI MCM PTM PRA & 16 & & \cmark & ~ & ~ & 1013 & ~ & ~ & ~ & ~ &\cmark& ~ & ~ & ~ & ~ & ~ & ~ & ~ & ~ & ~ & ~ \\ \hline
        \cite{song2019efficient} & - & OF UF TOD UEC TD RE & 6 & & \cmark &\cmark&\cmark& 50000 & ~ & ~ & ~ & ~ &\cmark& ~ & ~ & ~ & ~ & ~ & ~ & ~ & ~ & ~ & ~ \\ \hline
    \end{tabular}
    
}
 \caption{Comparing different detectors according to considered vulnerabilities and dataset used.}
 \label{tab:dataset}
\end{table*}

\section{Research Findings} \label{Sec:Findings}

\begin{table*}[t]
    \centering
    \fontsize{8}{8}\selectfont
    \setlength{\tabcolsep}{2pt} 
    \renewcommand{\arraystretch}{1.4} 
    \resizebox{\textwidth}{!}{
    \begin{tabular}{|r l| l l l l l l l|l l l l l l l l l l l l l l l|r r l l l|l|c|c|p{7em}|l|}
    \cline{3-29}
    \multicolumn{2}{c|}{}& \multicolumn{7}{c|}{Pre-processing}& \multicolumn{15}{c|}{ML model} & \multicolumn{5}{c|}{Dataset}\\

    \hline
    
        Ref & Name & \rot{90}{AST} & \rot{90}{CFG} & \rot{90}{word2vec} & \rot{90}{ngram} & \rot{90}{network} & \rot{90}{other} & \rot{90}{custom} & \rot{90}{LR} & \rot{90}{DT} & \rot{90}{SVM} & \rot{90}{RF} & \rot{90}{kNN} & \rot{90}{Ensamble} & \rot{90}{NN} & \rot{90}{CNN} & \rot{90}{RNN} & \rot{90}{LSTM/GRU} & \rot{90}{Transformer} & \rot{90}{GraphNN} & \rot{90}{Attention} & \rot{90}{Bidirectional} & \rot{90}{Other} & \rot{90}{\# Vulnerabilities} & \rot{90}{\# Contracts} & \rot{90}{SmartBugs Wild } & \rot{90}{Other} & \rot{90}{Custom} & \rot{90}{Requires Source} & \rot{90}{ExecTime} & \rot{90}{Avg F1-score} & Compares to ML & \rot{90}{Is Code Available? \ } \\ \hline
        
        \cite{cai2024fine} & - &\cmark &\cmark & ~ & ~ &\cmark & ~ & ~ & ~ & ~ & ~ & ~ & ~ & ~ &\cmark & ~ & ~ & ~ & ~ &\cmark & ~ & ~ & ~ & 7 & 6000 & ~ & ~ &\cmark & \cmark & n.a. & 90.57 & \cite{zhuang2020smart,wu2021peculiar,zhang2022cbgru} &\cmark \\ \hline
        \cite{ma2023hgat} & HGAT &\cmark &\cmark & ~ & ~ & ~ & ~ &\cmark & ~ & ~ & ~ & ~ & ~ & ~ & ~ & ~ & ~ & ~ & ~ &\cmark &\cmark & ~ & ~ & 4 & 7018 &\cmark & ~ & ~ & \cmark & 1.04 & 84.25 &  & ~ \\ \hline
        \cite{jain2023integrated} & - & ~ & ~ &\cmark & ~ &\cmark & ~ & ~ & ~ & ~ & ~ & ~ & ~ & ~ & ~ & ~ & ~ &\cmark &\cmark & ~ & ~ &\cmark & ~ & 13 & n.a. &\cmark &\cmark & ~ & & n.a & 96.5 & \cite{sun2021attention,tann2018towards, gogineni2020multi, momeni2019machine, liu2021combining, wang2020contractward, yu2021deescvhunter, jeon2021smartcondetect, zhuang2020smart} & ~ \\ \hline
        \cite{mi2023automated} & - & ~ &\cmark & ~ &\cmark & ~ &\cmark & ~ & ~ & ~ & ~ & ~ & ~ & ~ &\cmark & ~ & ~ & ~ & ~ & ~ & ~ & ~ & ~ & - & 169968 & ~ &\cmark &\cmark & & 3.27 & 97 &  \cite{gogineni2020multi, gao2019towards, lutz2021escort} & ~ \\ \hline
        \cite{cai2023combine} & Bi-GGNN &\cmark &\cmark & ~ & ~ & ~ &\cmark & ~ & ~ & ~ & ~ & ~ & ~ & ~ & ~ & ~ & ~ &\cmark & ~ &\cmark &\cmark &\cmark & ~ & - & 17458 &\cmark &\cmark & ~ & \cmark & n.a. & 91.1 & \cite{tann2018towards,qian2020towards} & ~ \\ \hline
        \cite{zhang2022smart} & MODNN & ~ & ~ & ~ & ~ &\cmark &\cmark &\cmark & ~ & ~ & ~ & ~ & ~ & ~ &\cmark & ~ & ~ & ~ & ~ & ~ & ~ & ~ & ~ & 12 & 18796 &\cmark &\cmark &\cmark & & n.a. & 94.8 &  \cite{ashizawa2021eth2vec, liao2019soliaudit, zhuang2020smart} & ~ \\ \hline
        \cite{zhang2022cbgru} & CBGRU & ~ & ~ &\cmark & ~ & ~ &\cmark & ~ & ~ & ~ & ~ & ~ & ~ & ~ & ~ &\cmark & ~ &\cmark & ~ & ~ & ~ &\cmark & ~ & 6 & >10000 &\cmark & ~ & ~ & \cmark & n.a. & 89.93 & \cite{yu2021deescvhunter, zhuang2020smart, wu2021peculiar, qian2020towards} &\cmark \\ \hline
        \cite{zhang2022spcbig} & SPCBIG-EC & ~ & ~ &\cmark & ~ & ~ & ~ &\cmark & ~ & ~ & ~ & ~ & ~ & ~ & ~ &\cmark & ~ &\cmark & ~ & ~ &\cmark &\cmark & ~ & 6 & >10000 & ~ & ~ &\cmark & \cmark & 9.8 & 90.79 & \cite{ashizawa2021eth2vec, zhuang2020smart, yu2021deescvhunter} &\cmark \\ \hline
        \cite{li2022detecting} & - & ~ & ~ & ~ &\cmark & ~ &\cmark &\cmark &\cmark & ~ &\cmark & ~ &\cmark & ~ & ~ & ~ & ~ & ~ & ~ & ~ & ~ & ~ & ~ & - & 3328 & ~ & ~ &\cmark & & n.a. & 75.3 & - & ~ \\ \hline
        \cite{xue2022xfuzz} & xFuzz &\cmark &\cmark &\cmark & ~ & ~ & ~ & ~ & ~ &\cmark & ~ & ~ & ~ &\cmark & ~ & ~ & ~ & ~ & ~ & ~ & ~ & ~ & ~ & 3 & 7391 & \cmark & ~ &\cmark & \cmark & 30 & - & - &\cmark \\ \hline
        \cite{hwang2022codenet} & CodeNet & ~ & ~ & ~ & ~ & ~ & ~ &\cmark & ~ & ~ & ~ & ~ & ~ & ~ & ~ &\cmark & ~ & ~ & ~ & ~ & ~ & ~ & ~ & 4 & 13443 &\cmark &\cmark & ~ & & 0.14 & 97.63 & - & ~ \\ \hline
        \cite{gu2022detecting} & CNN-BiLSTM & ~ & ~ & ~ & ~ & ~ &\cmark & ~ & ~ & ~ & ~ & ~ & ~ & ~ & ~ &\cmark & ~ &\cmark & ~ & ~ & ~ &\cmark & ~ & 5 & 3262 & ~ & ~ &\cmark & & n.a. & 83.63 & - & ~ \\ \hline
        \cite{ashizawa2021eth2vec} & Eth2Vec &\cmark & ~ & ~ & ~ & ~ & ~ & ~ & ~ & ~ & ~ & ~ & ~ & ~ & ~ & ~ & ~ & ~ & ~ & ~ & ~ & ~ &\cmark & 7 & 95152 & ~ & ~ &\cmark & \cmark & 0.371 & 57.5 & \cite{momeni2019machine} &\cmark \\ \hline
        \cite{jeon2021smartcondetect} & SmartConDetect & ~ & ~ & ~ & ~ & ~ &\cmark &\cmark & ~ & ~ & ~ & ~ & ~ & ~ & ~ & ~ & ~ & ~ &\cmark & ~ & ~ & ~ & ~ & 23 & 10000 &  & ~ &\cmark & \cmark & n.a. & 90.9 & \cite{ashizawa2021eth2vec,zhuang2020smart} & ~ \\ \hline
        \cite{sun2021attention} & ABCNN & ~ & ~ &\cmark & ~ & ~ &\cmark & ~ & ~ & ~ & ~ & ~ & ~ & ~ & ~ &\cmark & ~ & ~ & ~ & ~ &\cmark & ~ & ~ & 4 & 8632 & ~ &\cmark & ~ & & <1 & 87.66 & - & ~ \\ \hline
        \cite{yu2021deescvhunter} & DeeSCVHunter & ~ & ~ & ~ & ~ & ~ & ~ &\cmark & ~ & ~ & ~ & ~ & ~ & ~ & ~ &\cmark &\cmark &\cmark & ~ & ~ &\cmark &\cmark &\cmark & 2 & 40932 & ~ & \cmark & ~ & \cmark & n.a. & 83.4 & \cite{zhuang2020smart} &\cmark \\ \hline
        \cite{lutz2021escort} & ESCORT & ~ & ~ & ~ & ~ &\cmark & ~ & ~ & ~ & ~ & ~ & ~ & ~ & ~ &\cmark & ~ & ~ & ~ & ~ & ~ & ~ & ~ & ~ & 7 & 93497 & ~ & ~ &\cmark & & 0.20 & 95 & \cite{wang2020contractward, tann2018towards, gogineni2020multi,huang2018hunting} & ~ \\ \hline
        \cite{wu2021peculiar} & Peculiar &\cmark & ~ & ~ & ~ & ~ &\cmark & ~ & ~ & ~ & ~ & ~ & ~ & ~ & ~ & ~ & ~ & ~ &\cmark & ~ & ~ & ~ & ~ & 1 & 1648 &\cmark & ~ &\cmark & \cmark & n.a. & 92 & \cite{zhuang2020smart} &\cmark \\ \hline
        \cite{huang2021hunting} & - & ~ &\cmark & ~ & ~ & ~ & ~ & ~ & ~ & ~ & ~ & ~ & ~ & ~ & ~ & ~ & ~ & ~ & ~ &\cmark & ~ & ~ & ~ & 5 & >2M & ~ & ~ &\cmark & & 0.47 & n.a. & - & ~ \\ \hline
        \cite{liu2021combining} & CGE & ~ & ~ & ~ & ~ & ~ & ~ &\cmark & ~ & ~ & ~ & ~ & ~ & ~ &\cmark & ~ & ~ & ~ & ~ &\cmark & ~ & ~ & ~ & 3 & 40932 & ~ &\cmark & ~ & \cmark & n.a. & 85.43 & \cite{zhuang2020smart} & ~ \\ \hline
        \cite{zhuang2020smart} & TMP, DR-GCN & ~ &\cmark & ~ & ~ & ~ & ~ & ~ & ~ & ~ & ~ & ~ & ~ & ~ & ~ & ~ & ~ & ~ & ~ &\cmark & ~ & ~ &\cmark & 3 & 40932 & ~ &\cmark & ~ & \cmark & n.a. & 77.13 &  & ~ \\ \hline
        \cite{qian2020towards} & BLSTM-ATT & ~ & ~ &\cmark & ~ & ~ &\cmark & ~ & ~ & ~ & ~ & ~ & ~ & ~ & ~ & ~ & ~ &\cmark & ~ & ~ &\cmark &\cmark & ~ & 1 & 4000 & ~ & ~ &\cmark & \cmark & n.a. & 88.26 &  &\cmark \\ \hline
        \cite{wang2020contractward} & ContractWard & ~ & ~ & ~ &\cmark & ~ & ~ & ~ & ~ & ~ &\cmark &\cmark &\cmark &\cmark & ~ & ~ & ~ & ~ & ~ & ~ & ~ & ~ & ~ & 6 & 49502 & ~ & ~ &\cmark & & 4 & 97 & - & ~ \\ \hline
        \cite{liao2019soliaudit} & Soliaudit & ~ & ~ &\cmark &\cmark & ~ & ~ & ~ &\cmark &\cmark &\cmark &\cmark &\cmark & ~ & ~ &\cmark & ~ & ~ & ~ & ~ & ~ & ~ & ~ & 13 & 17979 & ~ & ~ &\cmark &  & n.a. & 90.4 & - &\cmark \\ \hline
        \cite{momeni2019machine} & - &\cmark & ~ & ~ & ~ & ~ &\cmark & ~ & ~ &\cmark &\cmark &\cmark & ~ & ~ &\cmark & ~ & ~ & ~ & ~ & ~ & ~ & ~ & ~ & 16 & 1013 & ~ & ~ &\cmark &\cmark & <0.001 & - & - & ~ \\ \hline
        \cite{song2019efficient} & - & ~ & ~ & ~ &\cmark & ~ & ~ & ~ & ~ & ~ &\cmark &\cmark &\cmark & ~ & ~ & ~ & ~ & ~ & ~ & ~ & ~ & ~ & ~ & 6 & 50000 & ~ & ~ &\cmark & & 4 & 93 & - & ~ \\ \hline

    \end{tabular}
    }
    \caption{Comparing different detectors according to pre-processing technique, models, datasetand results.}
    \label{tab:methods}
\end{table*}

After presenting the machine learning-based detectors we analyzed, we can finally address our research questions. The first two research questions we want to explore concern which vulnerabilities are the most considered (\textbf{RQ1}) and which datasets have been used to train the models (\textbf{RQ2}). We can answer both research questions by looking at Table \ref{tab:dataset}. Entries marked as "-" in the "\#" column for vulnerabilities indicate that either the detector functioned solely as a binary classifier or that the individual F1-scores for each class were not provided. First, we notice the heterogeneity of the datasets used for training the classifiers. The typical approach involves either considering a large number of vulnerabilities and proposing a binary classifier (vulnerable/not vulnerable) or focusing on a restricted number of vulnerabilities.

\begin{figure}[t]
            \resizebox{0.75\textwidth}{!}{%
    \includegraphics{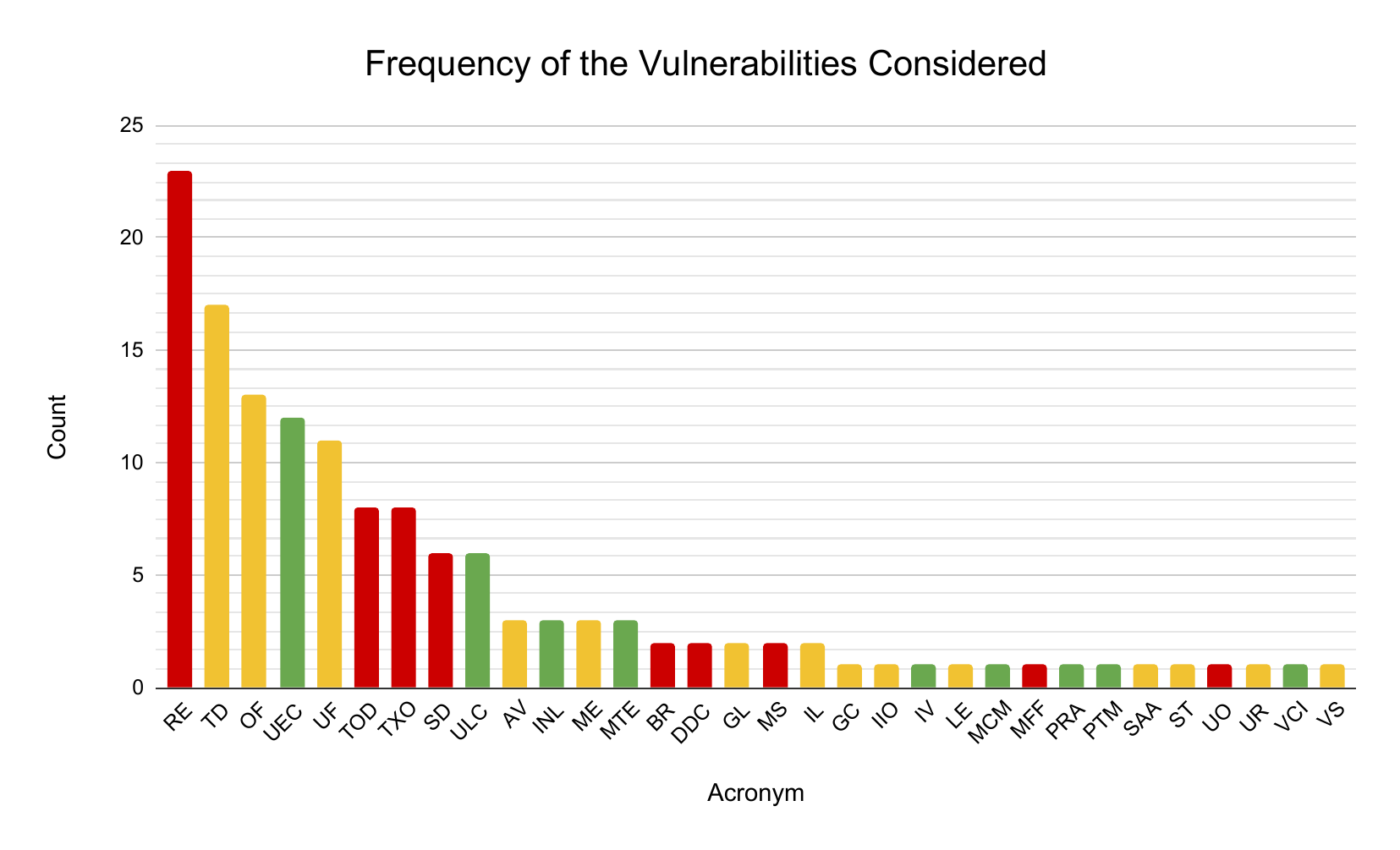}   } 
    \caption{Distribution of vulnerability searched by the detectors. }
    \label{fig:hist_vuln} 
\end{figure}

Figure \ref{fig:hist_vuln} shows the frequency with which certain vulnerabilities have been considered, addressing our \textbf{RQ1}. Reentrancy is the most frequently considered, appearing in all the papers we reviewed, followed by timestamp dependency and overflow issues. Most of the vulnerabilities in the tail of the histogram have been considered only by binary detectors. We found it particularly challenging to identify some of the vulnerabilities in certain papers, especially due to the lack or incompleteness of the definitions provided for each of the vulnerabilities considered. Most papers typically adopt the nomenclature used by the static analysis tools they deployed for the labeling process. 


For what concern \textbf{RQ2}, we observe that the most common solution for dataset construction is to create a custom dataset. Contracts are typically downloaded using software such as Etherscan or authors exploit the unlabeled, pre-cleaned Smartbugs Wild. The set of smart contracts, typically consisting of source code, is then automatically labeled using two or three static analysis tools. Labels are generally decided by majority voting; for example, if two out of three tools label the code as prone to a certain vulnerability, the contract is labeled accordingly. In these works, the authors usually do not include the definitions of the properties related to the vulnerabilities they are targeting or provide only a few restricted examples of code snippets. This is also the case when models are trained using multiple datasets from different sources (see the column \textit{\# contracts}, where we left separated the number of contracts in different datasets). As a result, the proposed detector is typically a binary classifier due to the heterogeneity and the cardinality of the vulnerabilities considered. Only a few authors focus on constructing datasets from scratch, and in these cases, we notice that their method to extract relevant features and structures from the data reflects their effort to understand which entities (i.e., variables, functions, external calls) are relevant for the task, which typically represents a significant part of their contributions.

Specifically, many works focus on the feature extraction step, which can be divided into different phases. First, the data is cleaned. In the case of source code, spaces, tabulations, and comments are removed. If the contracts are collected directly from Ethereum a large number of them is usually discarded, as they have the same checksum. Next, a reduction step occurs. For the source code, this might involve removing lines of code that are not relevant for detection, while for the opcode, it typically involves combining and/or removing similar operations, such as renaming PUSH1, PUSH2, etc., to PUSH. The final step is feature extraction, where a multitude of approaches have been applied, from constructing a graph representation to designing a model to learn the best features.
As far as \textbf{RQ2} is concerned, we can assert that these latter types of works achieve superior quality compared to others. The effort to detect relevant information for the subsequent detection step not only helps readers gain deeper insights into the problem but also provides a solid foundation for improving future work by including new vulnerabilities or new forms of the same vulnerability that might appear.

Other than the dataset, we are interested in analyzing which ML models have been used for the detection, which corresponds to our \textbf{RQ3}. To this purpose, we gathered all meaningful data in Table~\ref{tab:methods}. 
In this table we summarize the most important information concerning the data pre-processing method, the ML methods implemented, the number of smart contracts, as well as performance in terms of F1-score, execution time, and if the authors provided their dataset or implementation.
Except for a few cases, we can notice how most of the papers do not compare to related work. In these papers the authors decided to assess the superiority of their model with respect to vanilla networks, classic ML techniques or conducted an ablation study on the different components of the model, presenting how a specific combination of pre-processing techniques combined with the use of certain network achieves a higher detection rate.

Addressing \textbf{RQ4} has been the most complex part of our work. From the beginning, it was clear how the researchers addressed different aspects of the problem, for example by focusing on a specific architecture and its variations or by testing the best feature extraction methodology. Only a few works specifically defined the properties and the mechanisms necessary to have a vulnerability or an attack in terms of instructions, functions, and contracts. The complexity of the interactions and the possible operations of a contract with other ones makes it extremely difficult to identify all possible variations of a different vulnerability, especially when an attack can be performed thanks to two or more different vulnerabilities. While the average F1-score reported is usually over 90\% (see Table \ref{tab:methods}), we can see that the datasets used greatly vary, and when the authors replicate related works on their dataset obtain very different methods. 
The dataset labeling process is indeed the most critical aspect, given both the difficulty of the task and the high positive rates of static analyzers. Even manual labeling is not error-free as it is biased by the experience and opinion of the programmer. 
We talk in detail about this and other problems in Section \ref{sec:Comparison}.
Nevertheless, most authors do not provide the code or even the datasets they constructed. This issue raises many concerns about the genuineness of the results achieved.

\section{Limitations and Open Problems}\label{sec:Problems}
During our analysis of the literature, we uncovered several problems. To tackle these issues effectively, we have categorized them into four different groups. For each class, we also discuss the problem and propose specific mitigation strategies. Finally, we also present open problems that can inspire future work.

\subsection{Machine Learning-related}
ML and AI  typically have to deal with issues such as scalability, interpretability and replicability.


\begin{description}[leftmargin=1.5em]
    \item[\textbf{Scalability.}] A model's capacity to adapt to new tasks efficiently defines if it is scalable. In vulnerability detection, an ideal detector should rapidly adjust to identify emerging vulnerabilities. However, this is often impractical due to the time-consuming training required for the classifier to learn meaningful patterns. One potential solution involves training separate detectors for each vulnerability type, although this increases inference time. Notably, \cite{lutz2021escort} introduces a modular approach allowing for the detection of new vulnerabilities by adding branches to the model and leveraging transfer learning. Similarly, \cite{li2022detecting} and \cite{gu2022detecting} offer detectors with some degree of robustness to this issue. Unfortunately, their solutions are only able to distinguish between known or unknown vulnerabilities, making it difficult to understand where the problem is and how to address it.

    \item[\textbf{Interpretability.}] Explainability, or interpretability, has gained considerable attention in recent years, especially within the AI community \cite{arrieta2020explainable}. Explainable AI (XAI) is a specialized field aimed at shedding light on the inner workings of the traditional black-box paradigm, providing a better understanding of its predictions. Only one of the papers discussed in this SoK \cite{liu2021smart} explicitly addresses this aspect, focusing on the explainability of expert patterns and the integration of deep graph features. Attention mechanisms also play a role in enhancing model interpretability.

    \item[\textbf{Replicability.}] One of the major challenges in the AI community is replicability. Many researchers do not share the code used to evaluate their models or the datasets employed, particularly if they curated them manually. Training a model involves tuning numerous parameters, and attempting to replicate a method based solely on a paper's description is impractical, making it difficult to recreate the same environment and obtain comparable results.

\end{description}

These problems are well-known in the ML community, and some of them can be easily addressed. For instance, a model can be designed to be scalable by using modules that can be re-trained separately to accommodate new vulnerabilities. Other solutions include using fine-tuning to learn different forms of the same vulnerability or exploring anomaly detection techniques. Given the attention XAI has been receiving, many techniques are available to overcome the black-box model. Specifically, including the attention mechanism and carefully studying the feature extraction step not only allow the model to achieve better performance but could also make the functioning of the detector clear. Finally, the availability of the code and the dataset used are essential for replicability purposes, and in particular to enhance the transparency of the method used. 

While analyzing the detectors, we realized there are significant differences between the methods proposed by researchers coming from the machine learning community and those working on security. The first group often focuses on addressing these known issues and concentrates more on the design of the architecture, almost not caring about the nature of the data. On the other hand, researchers working on security usually present more thoroughly the vulnerabilities, but they typically do not compare their method to any other detectors and simply report the related work at the end of their paper. As a result, most of the works we examined either report very high performance in terms of F1-score, but they are shady about the data used and do not provide any explanation on the nature of the sample they failed to classify, or, in the case of papers published in security related conferences or journals, the authors typically present the problem very well, but the ML-based detector they train does not achieve comparable results.

\subsection{Comparison-related}
\label{sec:Comparison}
In order to prove the superiority of a method with respect to the others, it needs to be compared with related work. 

\begin{description}[leftmargin=1.5em]
    \item[\textbf{Lack of comparison with related work.}] 
    Papers published in conferences about security place the related work at the end (before the conclusion) and they do not worry about comparing with any of them. This is very different from what usually happens in pure ML/AI conferences, as to be published the authors need to demonstrate there are not any other methods in the literature like theirs or at least to compare with similar ones.

    \item[\textbf{Missing results for single vulnerability.}]
    Many of the examined papers only report the average F1-score, and not how the detector performs on a single vulnerability. 

    \item[\textbf{Using different metrics.}]
    Using multiple metrics is not a drawback, but reporting the F1-score should be a must. Some works, e.g., \citet{huang2021hunting,cai2024fine}, use other metrics such as precision, recall, or false positive rates. 

    \item[\textbf{Heterogeneity of used datasets.}]
    As can be seen from Figure \ref{fig:hist_det}, a common approach is to create a custom dataset. Unfortunately, many of these methods consider less than 10000 contracts and most of them do not treat underrepresented classes. Moreover, an even smaller group makes these datasets available. \new{Alternatively, many exploit either Smartbugs Wild, or the dataset provided from the Etherscan website, but they all apply different labeling methods.}

\begin{figure}[t]
    \centering
    \includegraphics[width=0.65\textwidth]{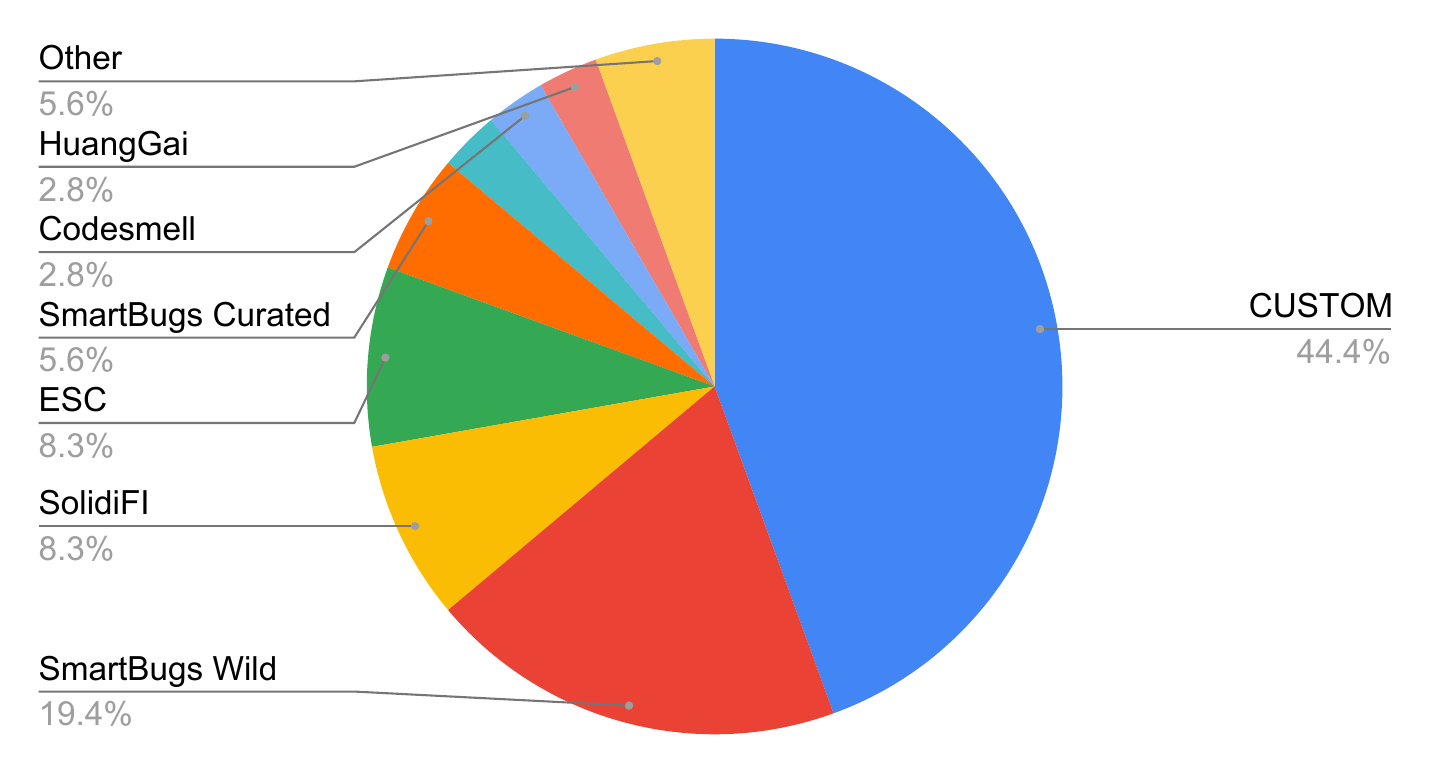}
    \caption{Percentage of methods using a custom dataset or an existing one.}
    \label{fig:hist_det}
\end{figure}

    \item[\textbf{Vulnerabilities nomenclature.}]
    Across the papers, the vulnerabilities are defined using different names, or they are grouped into macro categories that are not consistent with the others. Even if the work of Vidal et al. \cite{vidal2024openscv} greatly helped us in matching differently called vulnerabilities as being the same one, the authors themselves emphasize how difficult this process is due to ambiguous labels and unclear definitions. For example, one of the vulnerabilities authors detect in \cite{cai2024fine} defined as 'ERC20-TOD' refers to the definition of transaction order dependency provided in \cite{zhang2020framework}. In this paper,  Zhang et al. define a very specific case of TOD which depends on the presence of the 'approve' in a contract. As this is only a very specific case it would be difficult to compare the results with other methods detecting the same vulnerability.

    \item[\textbf{Labels granularity.}]
    The taxonomies adopted across the papers vary greatly. While some of the most common vulnerabilities share the same names, many labels are inconsistent due to differing nomenclatures and levels of detail. For instance, papers using the DASP top 10 taxonomy often have general classes like DoS and Access Control, which represent multiple vulnerabilities. Conversely, some authors distinguish between vulnerabilities within the same class. For example, \textit{Check effect} and \textit{Inline assembly} are treated separately in \cite{jain2023integrated}, though OpenSCV groups them as \textit{Improper Check of Low-Level Call Return Value}  (ULC according to Table \ref{tab:risks}). \citet{zhang2022smart} adopts the same differentiation, while also considering separately \textit{Low Level Calls} vulnerabilities. Neither of these papers unfortunately explain what properties are considered for each of these classes, or provide a code example. In a similar way, authors in  \cite{lutz2021escort} distinguish between \textit{Tainted self-distruct} and \textit{Accessible self-distruct}. In this case, however, it is simpler to separate the vulnerabilities: the first refers to the possibility of an external contract accessing the self-destruct functionality, while the second also involves an attacker changing the address parameter to transfer the balance to an address of their choice.

    \item[\textbf{Results inconsistency.}]
    Some papers replicate similar techniques with very different results from the ones reported originally. \citet{cai2024fine} propose a new dataset construction method based on Security Best Practices (SBP). They compare their method with \cite{wu2021peculiar,zhuang2020smart}. According to them, the methods achieve an F1-score of 56.45\% and 55.41\% respectively for reentrancy, contrary to the 92\% and 78.11\% declared in the papers. In this case, the results are highly dependent on the dataset, but it is hard to appoint one as the best, given the difference in labeling methods. This is not the only case, as other papers report different results when replicating similar methods~\cite{zhang2022cbgru}. 

\end{description}

During our study, we noticed how complicated it has been to compare the performance of different detectors. Once again, there is a notable difference between works according to the topic addressed by the journal or conference they published in. The comparison is made even more difficult due to the heterogeneity of methods, vulnerabilities, and datasets. While the creation of a benchmark dataset is of critical importance and might help with the comparison task, authors should always report at least the F1-score as a metric and differentiate the performance they achieved for each class of vulnerability. Moreover, even if taxonomies such as DASP10 and CWE are often chosen, their adoption should be discouraged, as the first one is too ambiguous \cite{vidal2024openscv} and the second one is no longer maintained since 2020.

\subsection{Usability-related} 

Vulnerability detectors should target usability as they need to be applicable in real-world scenarios.

\begin{description}[leftmargin=1.5em]
    \item[\textbf{Missing or high inference time.}]
    A proficient vulnerability detector must prioritize usability, being capable of efficiently identifying a broad spectrum of vulnerabilities within reasonable timeframes. Unfortunately, many methods fail to provide crucial information regarding the average inference time per contract. In some cases, detection processes may even exceed three seconds \cite{zhang2022spcbig, wang2020contractward, song2019efficient}.

    \item[\textbf{Vulnerability location.}]
    While most research focuses on detecting vulnerabilities at the contract level, few provide insights into specific vulnerability locations within the code. For instance, \cite{liao2019soliaudit} attempts to map vulnerability locations by feeding detection results back to the model, although the effectiveness of this approach remains uncertain. Conversely, \cite{cai2024fine} focuses solely on reentrancy detection. However, their detector's limitation to single lines of code restricts its applicability to vulnerabilities requiring analysis across multiple lines. Despite reentrancy often manifesting in a single line, its potential to spread across multiple lines cannot be overlooked.

    \item[\textbf{Source code vs. bytecode.}]
    Even if most of the examined detectors take the opcode of the smart contract for the detections, some of them require the source code instead. This approach limits the test cases to only the contracts where the source code is available. In the other case, instead, the detector might even work on EVM-compatible blockchains \cite{chen2020soda}.

    \item[\textbf{Risk of vulnerabilities.}]
    Figure \ref{fig:hist_vuln} shows how many detectors focus on specific vulnerabilities. Refer to Table \ref{tab:risks} for the meaning of the acronyms. The colors are also consistent with Table \ref{tab:risks} and represent the severity of the vulnerabilities. As can be seen, most of the detectors focus on reentrancy, but there are many other severe vulnerabilities (in red) that are rarely taken into consideration.

    \item[\textbf{Excluding contracts without vulnerabilities.}]
    Many of the papers overlook contracts without vulnerabilities, relying on datasets that include only contracts with multiple types of vulnerabilities \cite{ma2023hgat}. While it is useful to show the model's ability to distinguish between different vulnerabilities, this approach reduces the detector's practicality. As a result, the model will be unable to correctly identify a contract without vulnerabilities during inference, as it has never encountered one during training.
    
\end{description}




The design of a machine learning detector should target usability as its main purpose. This means providing a detector that is at least able to detect a wide range of vulnerabilities in real-time, as well as safe contracts. For this reason, many different classes of vulnerabilities should be considered in the dataset, while also designing a not-too-complex model which achieves low inference time. Unfortunately, this is not enough to make a detector really useful. From a user point of view, the best solution would be to also provide the location of such an issue and possibly propose a fix. Finally, when designing a detector, we should also keep in mind at which level we want to deploy such a tool. For example, if we are using the detector to find vulnerable lines of code at implementation time, it is different rather than using it to explore smart contracts already deployed, where most of the time we do not have the source code of the contract.

\subsection{Dataset-related}

The performance of a machine learning model always depends on the completeness and the correctness of the dataset they are trained on.

\begin{description}[leftmargin=1.5em]
    \item[\textbf{Limited number of contracts.}] 
     A third of the methods presented exploit a dataset consisting of less than 10000 smart contracts. While this might still be acceptable for detectors focusing on one single vulnerability, the number is too shallow for most of the detectors, especially considering the extremely high number of identical contracts. 

    \item[\textbf{Untreated unbalanced classes.}]
    Only a few works treat the problem. \new{The method presented in} \cite{wang2020contractward}  reports an F1-score of over 96\% despite the average 4 seconds required to classify a single contract, while \new{authors in }\cite{lutz2021escort} creates its custom dataset from over a million of contracts allocating at least 15000 contracts for each class. \new{Authors in} \cite{jeon2021smartcondetect} report using 10000 contracts to detect 23 different types of vulnerability, showing the area under the curve (AUC) as a metric. Considering how the AUC reaches 100\% for more than one vulnerability, it is clear how those classes have been highly under-represented, invalidating most of their results.

    \item[\textbf{Unclear dataset creation process.}]
    \new{Two of the papers examined }\cite{gu2022detecting, li2022detecting} propose novel methods for vulnerability detection, leveraging the SODA framework \cite{chen2020soda} and referring to 'Geth instrumentation'/'Geth instrumentation collections' without providing adequate explanation or reliable references. \new{In other papers} \cite{zhuang2020smart, liu2021combining, yu2021deescvhunter} \new{the authors} utilize the \textit{ESC} dataset, comprising over 40,000 contracts, \new{with only one of them }offering access to the dataset's code and data \cite{yu2021deescvhunter}. \new{Moreover, in this last one the authors }
    focus solely on two out of the three vulnerabilities in the dataset without specifying the number of contracts considered. \new{In a similar way, another work}   detects reentrancy on SmartBugs Wild without detailing the number of contracts containing this vulnerability \cite{wu2021peculiar}. \new{Finally, authors in} \cite{cai2023combine} provide a broken link in the dataset specifications. 

    \item[\textbf{Labeling methods.}]
    While some authors labeled the dataset manually, the most common approach is to exploit the methods presented in Section \ref{sec:formal}. Unfortunately, the current static analyzers available to create the training datasets are hardly accompanied by formal soundness guarantees (with few exceptions, notably \new{the work proposed in} \cite{schneidewind2020good}). This limitation suggests that machine learning algorithms currently in use may have been trained on mislabeled observations. A recent study evaluated the performance of 9 static analysis tools on different datasets, including SmartBugs Wild, achieving notably low accuracies \cite{durieux2020empirical}. \new{Even the same authors who created Smartbugs, recently highlighted two major problems related to the dataset. The first issue concerns how a specific vulnerability is defined: different experts can label the same smart contract with opposite labels, according to which property they deem important for the labeling \cite{di2023consolidation}. The second problem instead is related to maintenance: as the compiler version changes or new vulnerabilities are discovered, a large number of static analyzers exhibit growing error and failure rates \cite{di2024evolution}. 
    Finally, the distinction between contract files and contracts is shady. While some papers adopt \cite{durieux2020empirical} approach and label the whole \textit{.sol} file, others label the single contracts within it. Unfortunately, most of the papers are not transparent about their labeling method, further complicating the analysis by making unreliable the number of contracts reported in their dataset.}

\end{description}


The selection of an appropriate dataset is a critical issue. Although Ethereum is a public blockchain with millions of blocks, only a small fraction of stored contracts have accessible source code. 
This is because only the bytecode is required to be stored on the blockchain, not the source code. A contract's code is accessible only if the developer explicitly includes it during deployment. Furthermore, many of these contracts are identical, which significantly reduces the number of unique contracts available. 
In this context, Smartbugs Wild represents the most viable alternative, being the largest dataset of unique Ethereum smart contracts available, with about 47,000 contracts. Despite this extensive collection, two major issues persist: labeling and class imbalance. Significant efforts are still needed to address these issues. Currently, a combination of manual labeling and static analyzers represents the state-of-the-art approach, with careful consideration of their limitations.

When selecting static analysis tools, it is important to consider which vulnerabilities they are designed to detect and whether they are actively maintained. Simultaneously, multiple individuals should perform manual labeling to mitigate potential bias. Additionally, the number of examples for each class of vulnerability varies greatly. For the most underrepresented classes, it may be beneficial to increase the number of examples using data augmentation techniques, such as oversampling, or creating new contracts through bug injection.

\subsection{Open Problems}
\label{Sec:Open}

While static analysis presents as a robust and versatile method, it is tailored to specific security properties dictated by the safety standards it aims to uphold. Consequently, while these tools effectively identify particular vulnerabilities, their scope remains constrained. Combining various analyzers may mitigate this limitation, yet applying multiple analyses to the same code can incur significant computational costs and pose the risk of inconsistent outcomes \cite{momeni2019machine}. Although techniques like taint analysis and symbolic execution are theoretically sound, recent research demonstrates that current static analyzers often fall short, producing false negatives \cite{schneidewind2020ethor,schneidewind2020good, zhang2020framework}.
Machine Learning offers a viable alternative. ML frameworks can harness static analysis by utilizing multiple analyzers to generate diverse labeled datasets, enabling the training of ML models for classification and vulnerability detection. Several ML experiments have been documented in recent literature.

An additional concern for existing ML frameworks is their precision. Establishing a standardized dataset to assess the efficacy of various approaches would be invaluable, aiding in selecting the most effective tool. However, this remains a formidable challenge due to the significant variations in dataset size, code format (source code vs bytecode vs opcode), and ML techniques employed for analysis \cite{chakraborty2021deep}.
Despite numerous detectors proposed in the literature, based on formal methods and machine learning, this survey highlights two main open problems. The first and most evident issue is the lack of a benchmark dataset. While Smartbugs Wild offers a large collection of unique contracts, the authors themselves point out the numerous issues with currently available tools for labeling. This concern is compounded by a second, less obvious problem: the ambiguity in defining the properties and conditions that determine the presence or absence of specific vulnerabilities. 

Some vulnerabilities, like overflow issues, poor randomness, and unchecked low-level calls, are straightforward to label. However, others are highly dependent on the smart contract's semantics. For instance, if a smart contract aims to provide a certain level of user access, defining an access control vulnerability without prior knowledge of the contract's purpose is challenging. 
Moreover, certain scenarios require reliance on external factors, such as block timestamps, to activate specific smart contract functions (e.g., unlocking a certain amount of ether or executing repeated payments over time). In these cases, identifying vulnerabilities is challenging because it requires evaluating whether an attacker’s manipulation of the timestamp would significantly impact the contract’s behavior (e.g., determining a lottery winner) or if minor fluctuations are negligible (e.g., in the case of repeated payments or expiration times).

\medskip
Another significant issue is the constant evolution of vulnerability properties due to compiler updates and the introduction of new functions and keywords in the Solidity language. For example, starting from version 0.8.0, the compiler includes automatic checks for underflow and overflow issues. Programmers can bypass these checks to save gas using the \lstinline{unchecked} keyword. This change divides contracts into two categories: those released before version 0.8.0, which require checks for arithmetic issues, and those released after, which need checks only when the \lstinline{unchecked} keyword is used. 
Ultimately, there are vulnerabilities that have never been considered because it is difficult to account for all best practices and new features or language constructs added to Solidity. For example, the use of \lstinline{send} and \lstinline{transfer} should always be discouraged, and \lstinline{call} should be used instead. The \lstinline{send} and \lstinline{transfer} functions impose a gas limit of 2300, which is designed to prevent reentrancy attacks by limiting the amount of gas available to the receiving contract. However, this gas limit can also cause transactions to fail if the receiving contract requires more gas to execute its fallback function. Moreover, after the Istanbul hard fork, gas costs have increased, making the fixed gas limit of 2300 too low for some operations that were previously possible. For this reason, \lstinline{call} with appropriate safety checks (e.g., checking the return value and limiting gas) is more flexible and provides more control and security. 
Although this is the recommendation of the community\footnote{https://solidity-by-example.org/sending-ether/} up to date, \citet{wohrer2018smart} suggested avoiding the use of \lstinline{call} for eliminating the risk of reentrancy. This proves how it is difficult to define best practices in the long run that are consistent. 

Overall, addressing these challenges is crucial for advancing the reliability and accuracy of vulnerability detection in smart contracts. Future research should focus on creating standardized benchmark datasets and refining definitions of vulnerabilities to improve the consistency and effectiveness of both static analysis and machine learning approaches. By doing so, the field can move towards more robust and comprehensive security solutions for smart contracts.

\section{Conclusion}
\label{sec:Conclusion}

In this survey we presented an in-depth analysis of machine learning-based methods for vulnerability detection in Ethereum smart contracts. Even if intelligent detectors hold the potential to detect a wide range of vulnerabilities in less than one second per contract, current literature is held back by several issues. The lack of a robust benchmark dataset while the results are shadowed by the unclear description of the techniques and the dataset used, as well as by the difficulty of comparing different detectors. Despite existing open challenges, our proposed guidelines offer essential strategies to address some of the most pressing issues in the domain.


\section*{Acknowledgements}
This study was carried out within the PE0000014 - Security and Rights in the
CyberSpace (SERICS) and received funding from the European Union Next-GenerationEU - National Recovery and Resilience Plan (NRRP) – MISSION 4
COMPONENT 2, INVESTIMENT 1.3 – CUP N. H73C22000890001. This work
has been also partially supported by the Research Project INDAM GNCS 2024
- CUP E53C23001670001 “Modelli composizionali per l’analisi di sistemi reversivili distribuiti (MARVEL)” and by the Project PRIN 2020 - CUP N. 20202FCJMH "NiRvAna - Noninterference and Reversibility Analysis in Private Blockchains". This manuscript
reflects only the authors’ views and opinions, neither the European Union nor
the European Commission can be considered responsible for them.

\bibliographystyle{ACM-Reference-Format}
\bibliography{biblio}

\newpage
\appendix

\section*{Appendix} \label{Appendix}

We present a selection of source code examples highlighting common vulnerabilities in Solidity smart contracts. 
Each example corresponds to a category defined in Section \ref{sec:vulnerabilities}, focusing on the most significant vulnerability within each category.

\new{\subsection*{Reentrancy}}
The first snippet shows a typical reentrancy scenario involving two distinct contracts: the \lstinline{msg.sender.call("")} invocation passes an empty string as an argument, leading to the invocation of the fallback function in the callee code.
The problem arises because \lstinline{balances[msg.sender]} is set to 0 \emph{after} the call: if an attacker is exploiting the reentrancy, the \lstinline{withdraw} function is invoked over and over again and the balance keeps being greater than 0, eventually emptying the whole wallet.
Proposing the following code example, we assume the attacker deposited 1 ether in a previous interaction. 
\begin{lstlisting}[language=Solidity]
// UNSAFE EXTERNAL CALL: Reentrancy

// CONTRACT 1
contract EtherStore {
  mapping(address => uint) public balances;
  
  function deposit() public payable {
    balances[msg.sender] += msg.value;
  }
  
  // VULNERABLE FUNCTION
  function withdraw() external {
    uint256 amount = balances[msg.sender];
    (bool success,) = msg.sender.call{value: balances[msg.sender]}("");
    require(success);
    balances[msg.sender] = 0;
  }
}

// CONTRACT 2
contract Attacker {
  fallback() external payable {
    if (address(msg.sender).balance >= 1 ether) {
      EtherStore(msg.sender).withdraw();
    }
  }
}
\end{lstlisting}

\begin{lstlisting}[language=Solidity]

\end{lstlisting}

\new{\subsection*{Denial Of Service}}

In a simplified auction contract, the previous bidder is refunded every time a new bid is made. 
However, this system is vulnerable to attacks by malicious actors who can prevent other bidders from winning the auction. 
This can be done by blocking any incoming refund through a new bid.
The attacker can write a malicious contract that makes the send call fail by reverting any payment in the fallback function. 
This can effectively prevent any other bidder from winning the auction.

\begin{lstlisting}[language=Solidity]
// MISHANDLED EVENT: DoS
function bid() payable {
  require(msg.value > highestBid);
  require(currentLeader.send(highestBid));
  currentLeader = msg.sender;
  highestBid = msg.value;
}
\end{lstlisting}

\new{\subsection*{Multiple Sends}}

In the following code snippet, the repeated invocation of the \lstinline{send} method within a loop can lead to rapid gas depletion. This occurs because each \lstinline{send} operation within the same contract consumes a significant amount of gas, especially when executed in a loop, as illustrated below.
\begin{lstlisting}[language=Solidity]
// GAS DEPLETION: multiple sends
Payee[] payees;
function payOut() {
  uint256 i = 0;
  while (i < payees.length) {
    payees[i].addr.send(payees[i].value);
    i++;
  }
}
\end{lstlisting}

\new{\subsection*{Variable Shadowing}}

Local variable shadowing can be overlooked and lead to unwanted behaviors. Below, we illustrate a scenario demonstrating how this type of vulnerability can happen.
\begin{lstlisting}[language=Solidity]
// BAD PROGRAMMING: variable shadowing
contract SuperContract {
  uint a = 1;
}

contract SubContract is SuperContract {
  uint a = 2;
}
\end{lstlisting}

\new{\subsection*{Transaction Order Dependency}}

In the following example, the smart contract enables users to withdraw a specified amount of Ether only if they provide a correct solution, which hash matches the hash specified in the \lstinline{secret} variable. This smart contract is vulnerable to a transaction dependence vulnerability because an attacker can observe pending transactions in the Ethereum mempool, extract the correct solution submitted by a legitimate user, and then submit a new transaction with the same solution but with a higher gas price. This strategy could allow the attacker's transaction to be processed first and preemptively withdraw the Ether meant for the original user, effectively diverting the funds.
\begin{lstlisting}[language=Solidity]
// INCORRECT CONTROL FLOW: Transaction Ordering Dependence
bytes32 constant secret;
uint amount = 100 ether;

function withdraw(string solution) public {
  require(secret == sha3(solution));
  msg.sender.transfer(amount)
}
\end{lstlisting}

\new{\subsection*{Integer Overflow}}

The next example is taken from a smart contract that allows users to withdraw a specified amount of funds, but only after a predetermined timelock period has elapsed. Additionally, it offers a feature through the \lstinline{increaseLockTime} function, allowing users to voluntarily extend the waiting period for enhanced security or other reasons. However, this contract is vulnerable to an overflow attack. An attacker might exploit the \lstinline{increaseLockTime} function to manipulate the contract’s logic, inducing an overflow condition. This could prematurely reset the timelock, potentially enabling the attacker to execute a withdrawal before the intended expiration of the original timelock, thereby compromising the contract's security mechanism.
This type of vulnerability has been solved by Solidity 0.8, which now throws errors on overflow/underflow. 
\begin{lstlisting}[language=Solidity]
// ARITHMETIC ISSUES: overflow
function increaseLockTime(uint secs) public {
  lockTime[msg.sender] += secs;
}
\end{lstlisting}

\new{\subsection*{Tx.origin}}

In the last example, we analyze a withdraw function that allows the owner of a Wallet contract to withdraw a specified amount of funds. In this scenario, an attacker can craft a deceptive transaction that, when executed by the contract owner, inadvertently triggers the withdrawal function.  The crux of this vulnerability lies in the misuse of \lstinline{tx.origin} for authentication, which checks the original transaction sender's address instead of the immediate caller's address.
\begin{lstlisting}[language=Solidity]
// IMPROPER ACCESS CONTROL: Tx.origin 

contract Wallet {
  address owner;
  [...]
  function withdraw(uint amount, address to) public {
    require(tx.origin == owner);
    to.transfer(amount);
  }
}

contract Attacker {
  function attack(uint amount, walletAddress) {
    Wallet(walletAddress).withdraw(amount, attackerAddress);
  }
}
\end{lstlisting}

\end{document}